\documentstyle[aps,prb,preprint,epsf]{revtex}
\begin{document}
\title{Influence of typical environments on quantum processes}
\author{Ping Ao, Staffan Grundberg and J{\o}rgen Rammer}
\address{Department of Theoretical Physics, Ume{\aa} University, S-901
  87 UME{\AA}, SWEDEN}
\date{10 January 1996}
\maketitle
\begin{abstract}
  We present the results of studying the influence of different
  environmental states on the coherence of quantum processes. We
  choose to discuss a simple model which describe two electronic
  reservoirs connected through tunneling via a resonant state.  The
  model could, e.g., serve as an idealization of inelastic resonant
  tunneling through a double barrier structure. We develop Schwinger's
  closed time path formulation of non-equilibrium quantum statistical
  mechanics, and show that the influence of the environment on a
  coherent quantum process can be described by the value of a
  generating functional at a specific force value, thereby allowing
  for a unified discussion of destruction of phase coherence by
  various environmental states: thermal state, classical noise, time
  dependent classical field, and a coherent state. The model allows an
  extensive discussion of the influence of dissipation on the coherent
  quantum process, and expressions for the transmission coefficient
  are obtained in the possible limits.
\end{abstract}

\section{Introduction}

The progress in fabrication of submicron structures has lead to a
wealth of new structures whose transport properties are dominated by
the feature that electronic transport through the structure takes
place coherently. However, additional degrees of freedom in addition
to those of single electrons are present and it is of importance to
account for their influence on coherent processes. In the following we
shall investigate how different environmental states influence the
quantum process of main interest.  Although we shall choose a model
which is directly relevant to transport properties of a double barrier
structure or transport through a quantum dot, the technique we develop
is of general interest.

The paper is organized as follows: In section \ref{sec:model} we model
the system whose transport properties will be studied, and in section
\ref{sec:trans} we show how to describe the transmission probability
of a structure in terms of Green's functions.  In section
\ref{sec:csd} we discuss the main approximation, which will allow an
analytical description of the influence of dissipation on the
transmission properties. In section \ref{sec:path} we introduce the
closed time path formulation and develop the description of the
destruction of phase coherence in terms of a generating functional
technique, and in section \ref{sec:environments} we discuss the
effects of different environmental states: thermal and coherent
states, the external field case, and the fluctuating level model. We
shall for all these cases derive closed expressions for the
transmission probabilities, and compare the results in order to notice
similarities and differences in the influence of different
environments. In section \ref{sec:destruction} we discuss destruction
of phase coherence using the generic Aharonov-Bohm situation.  Finally
in section \ref{sec:summary} we summarize and conclude.

\section{The model for studying destruction of phase coherence}
\label{sec:model}

In this section we shall set up a minimal model describing the
dissipative feature of the general quantum transport problem. The
simplifications are introduced in order to be able to treat
analytically the influence of environments on the transport process.
We shall study the general problem of quantum transport where
electronic current reservoirs provide electrons to an active region
where interaction with an environment can take place. The reservoirs
corresponds in reality to large electrodes, and we can describe the
Hamiltonians for the left and right electrodes in terms of their
electron energy levels
\begin{equation}
  H_{\sigma}=\sum_{\bf p} \epsilon_{{\bf p},\sigma}
             a_{{\bf p},\sigma}^{\dag} a_{{\bf p},\sigma}.
\end{equation}
The quantum number ${\bf p}$ labels the momentum of the electron
states, and $\sigma=l,r$ refers to the left and right electrode,
respectively, and $a_{{\bf p},{\sigma}}^{\dag}$ creates particles in
states with these quantum numbers with the corresponding energy
${\epsilon}_{{\bf p},{\sigma}}$.

The central active (sample) region can in the absence of coupling to
the reservoirs be described by a Hamiltonian or equivalently its
energy levels.  The single particle energy at the central site,
labeled $c$, is $\epsilon_c$ corresponding to the term in the
Hamiltonian
\begin{equation}
  H_{s}=\epsilon_c a_c^{\dag} a_c.
\end{equation}

In the following we shall consider a model which allows extensive
analytical calculations, and therefore restrict the number of levels
relevant in the central region to one. Eventually we shall consider
also the case of two levels.  In the event that the reservoirs are
connected through the central site, electrons are transmitted between
them. Such a situation can quite generally be modeled by transfer
matrix elements,$V_{{\bf p},\sigma}$, between the reservoirs and the
central region.  The coupling of the electrodes to the central site
is therefore described by the tunneling Hamiltonian \cite{Bardeen}
\begin{equation}
  H_t = \sum_{{\bf p}\sigma} \{ V_{{\bf p}\sigma} a_c^{\dagger}
  a_{{\bf p}\sigma} + {\rm h.c.} \}.
\end{equation}
The transfer matrix elements are here considered to be
phenomenological parameters, but can of course for any chosen
microscopic model of say a double barrier, be expressed in terms of
the potential profile and the carrier mass. The Hamiltonian, $H_{e}$,
for the electronic system of interest is therefore given by
\begin{equation}
  H_e=H_l+H_r+H_s+H_t.
\end{equation}

Within the sample region we allow for interaction with an
environment.  For our purpose we can quite generally assume a bosonic
environment with a corresponding Hamiltonian $H_{b}$ which has the
standard normal mode form
\begin{equation}
  H_b = \sum_{\alpha} \hbar \omega_{\alpha} \left\{
  b_{\alpha}^{\dag} b_{\alpha} + \frac{1}{2} \right\}
\end{equation}
in terms of the bosonic creation and annihilation operators,
$b_{\alpha}^{\dag}$ and $b_{\alpha}$.

For the coupling of an electron at the central site we take a linear
coupling to the normal modes
\begin{equation}
  H_i=a_c^{\dag}a_c X,
\end{equation}
where $X$ is the collective environment displacement operator
\begin{equation}
  X = \sum_{\alpha} \lambda_{\alpha}
  \left\{ b_{\alpha}^{\dag}+b_{\alpha} \right\},
\end{equation}
and $\lambda_{\alpha}$ the coupling constant to mode $\alpha$.

The resulting Hamiltonian for system and environment
\begin{equation}
  H= H_e+H_b+H_i
\end{equation}
has been discussed in a variety of contexts \cite{Anderson}, most
recently in the context of inelastic resonant tunneling
\cite{Glazman,Wilkins,Resonant}, where the model has served as a
simplified description of the influence of interaction with phonons on
the transport properties of a double barrier structure. In the present
account we shall not only discuss a thermal environment, but a variety
of environmental states and their influence.  A purpose of the paper
is to present a calculational scheme that allow a unified description
of arbitrary environments, and in a direct way exhibits the physical
content of a model so that calculations within more realistic models
can be made tractable.  We now proceed to describe in detail the
transmission properties of the system under consideration.

\section{Transmission properties}
\label{sec:trans}

In mesoscopic physics where the main feature of electronic transport
is its coherence, the transport description can be expressed in terms
of the scattering properties of the mesoscopic structure. The choice
of model has been dictated by this feature, and in the following we
shall study the quantum mechanical problem of transmission of an
electron through a region where it can interact with additional
degrees of freedom. We wish, therefore, to calculate the transmission
coefficient for a particle emitted, say, from the left reservoir under
the assumption that it propagates via the central site, where it is
allowed to interact with an environment, to the right reservoir.  In
accordance with the initial condition of an electron impinging from
the left reservoir, we can assume that at some initial time $t_{i}$
the two subsystems, particle and environment, are decoupled, so that
the initial state is described by a separable statistical operator
\begin{equation}
  \rho_i = P_{{\bf p}',l} * \rho_b,
\end{equation}
where $P_{{\bf p}',l}$ is the projection operator describing an
electron in the left electrode in momentum state ${\bf p}'$
\begin{equation}
  P_{{\bf p}',l} = | {\bf p}',l \rangle \langle {\bf p}',l |.
\end{equation}
The state of the environment is presently arbitrary and described by
the statistical operator $\rho_{b}$. As regards the transmission
problem, given the above initial state, all information can be
extracted from the probability, $P_{{{\mbox{\scriptsize\bf p}}'},
  l\rightarrow {\bf p},r}(t)$, to find the electron in the right
electrode at time t with momentum ${\bf p}$. This conditional
probability is given by the expression
\begin{equation}
  \label{eq:condprob}
   P_{{\bf p}',l \rightarrow {\bf p},r}(t) =
  {\rm Tr} (\rho_iU^{\dagger}(t,t_i)P_{{\bf p},r}U(t,t_i)),
\end{equation}
where $P_{{\bf p},r}$ is the projection operator
\begin{equation}
  P_{{\bf p},r}=|{\bf p},r\rangle\langle {\bf p},r|
\end{equation}
corresponding to the assumed final outgoing particle state with
momentum ${\bf p}$ in the right electrode, and
\begin{equation}
  U(t,t_{i})= e^{-\frac{i}{\hbar}H(t-t_i)}
\end{equation}
is the evolution operator corresponding to the total Hamiltonian. The
trace with respect to all the degrees of freedom is denoted by Tr. The
absence of any environment operator discriminating the final states of
the environment is in accordance with the typical experimental
condition pertaining to the electronic conduction process, namely that
the environmental degrees of freedom are not observed.
                 
For an electron to propagate between the reservoirs it first has to
enter the sample region and lastly to exit it. In the event we only
explicitly consider interaction with the environment in the sample
region, we can exploit this feature and introduce a discussion in
terms of the effective coupling between reservoirs and sample region.
The chosen model has this feature to the extreme.  In the assumed
model the electron only couples to the environment at the central
site. We can therefore express the transition probability in a form
that explicitly only involves the dynamics of the electron at the
central site and the environment by simply noting, that in order to
calculate the amplitude for a transition from the left to the right
reservoir, the first propagation has to be from the left reservoir to
the central site, and the last propagation from the central site to
the right reservoir.  The transition probability can therefore be
rewritten as
\begin{eqnarray}
  \label{eq:transprob}
  P_{{\bf p}',l \rightarrow {\bf p},r}(t) &=& \frac{1}{\hbar^4}
  |V_{{\bf p}r}|^2 |V_{{\bf p}'l}|^2\int_0^t \!\! dt_1\,
  \int_0^t \!\! dt_2\, \int_0^t \!\! dt_3\, \int_0^t
  \!\! dt_4 \nonumber \\
  && \exp \left\{ \frac{i}{\hbar}\epsilon_{{\bf
        p},r}(t_2-t_3)+\frac{i}{\hbar}\epsilon_{{\bf p}',l}(t_4-t_1)
  \right\} \nonumber \\
  &&\langle  \hat{G}^A_c(t_4,t_3) \hat{G}^R_c(t_2,t_1) \rangle.
\end{eqnarray}
where $\langle ...\rangle=tr(\rho_b...)$ is shorthand for the trace with
respect to the environmental degrees of freedom, weighted with respect
to the initial environment state, and
\begin{equation}
  \hat{G}_c^R(t,t')=-i\theta(t-t')\langle 0|
  [a_c(t),a_c^{\dag}(t')]| 0\rangle
\end{equation}
is the retarded Green's operator (operator with respect to the
environmental degrees of freedom) for the central site dynamics as
\begin{equation}
  a_c(t)=\ e^{\frac{i}{\hbar}Ht} a_c e^{-\frac{i}{\hbar}Ht},
\end{equation}
and $|0\rangle$ denotes the particle vacuum state. The advanced Green's
operator is given by hermitian conjugation
\begin{equation}
  \label{eq:hermconj}
  \hat{G}_c^A(t,t')=[\hat{G}_c^R(t',t)]^{\dag},
\end{equation}
where ${\dag}$ denotes hermitian conjugation. In
eq.(\ref{eq:transprob}), the arbitrary initial time has been chosen at
time zero, $t_{i}=0$.

In many cases of interest we do not need the full information on the
transition probability as function of time. For instance, if we are
only interested in the average particle flow in a steady state we are
only interested in the transition probability per unit time
\begin{equation}
  \label{eq:tprobpertime}
    w_{{\bf p}',l \rightarrow {\bf p},r} = \lim_{t \rightarrow
    \infty} \frac{P_{{\bf p}',l \rightarrow {\bf p},r}(t)}t,
\end{equation}
or the transmission coefficient for making a transition between a
state of energy $\epsilon '$ in the left reservoir and a state of
energy $\epsilon$ in the right reservoir
\begin{equation}
  \label{eq:transfirst}
  T(\epsilon,\epsilon') = h \sum_{{\bf p}{\bf p}'} w_{{\bf p}',l
    \rightarrow {\bf p},r} \delta(\epsilon-\epsilon_{{\bf p},r})
  \delta(\epsilon'-\epsilon_{{\bf p}'l}).
\end{equation}
Invoking the scattering approach for the description of transport
properties for coherent quantum processes \cite{Landauer} we have at
zero temperature the contribution to the conductance at energy
$\epsilon'$
\begin{equation}
  G(\epsilon') = \frac{2e^2}{h} T(\epsilon'),
\end{equation}
where
\begin{equation}
  \label{eq:tot_trans_coeff}
  T(\epsilon ' ) = \int_0^{\infty}\!\!d\epsilon \,
  T(\epsilon,\epsilon'),
\end{equation}
is the total transmission coefficient, the probability to reach the
right reservoir for a state of energy $\epsilon'$ in the left
electrode. The factor of two is the spin degeneracy factor of the
electron.
 
\section{The wide band approximation}
\label{sec:csd}

We have in the previous section reduced the expression for the
transition probability to an expression which only involves the
dynamics at the central site. However, since both tunneling and
interaction with the environment is present the dynamics is
complicated and no closed expression for the transmission probability
can be found, that is, without any further assumption no
simplification is possible as the phonon average cannot be done
explicitly.  If we, however, assume that the widths, $W$, of the
electronic energy bands of the electrodes are the largest energy in
the problem, $W > \hbar\omega_{\alpha}$, it is possible to obtain
closed expressions for the transition probability for various cases of
environmental states. In this wide band limit an electron at the
central site decays into a continuum of states in the electrodes, and
we expect to have a quasi-stationary state which decays exponentially
in time. To utilize this property of the electrodes we first discuss
the lifetime for occupation of the central site in the absence of an
environment.

In the absence of coupling to the environment the retarded Green's
operator for the central site, $\hat{G}_c^R$, reduces to the c-number
Green's function, equaling the amplitude $G_c^R$ for the particle to
remain at the central site. The Fourier transformed amplitude
satisfies the Dyson equation
\begin{equation}
  G^R_c(\epsilon)=g^R_c(\epsilon)+g^R_c(\epsilon)
  \Sigma^R_c(\epsilon)G^R_c(\epsilon),
\end{equation}
where the central site self-energy is given by
\begin{equation}
  \Sigma^R_c(\epsilon) = \sum_{{\bf p}\sigma}|V_{{\bf p}\sigma}|^2
  g^R_{{\bf p}\sigma}(\epsilon),
\end{equation}
and describes the coupling of the central site to the reservoirs, as
indicated by the appearance of the electrode propagators.  A lower
capital is used to designate a subsystem Green's function, implying
that the electron is not allowed to propagate in or out of the
electrodes, i.e., its dynamics is determined by the Hamiltonian with
all connecting elements, $V_{{\bf p},{\sigma}}$, set equal to zero.

In the absence of tunneling between the electrodes and the central
site we have for the isolated electrode Green's functions
\begin{equation}
  g^R_{{\bf p}\sigma}(\epsilon)=
  \frac{1}{\epsilon-\epsilon _{{\bf p},\sigma}+i0},
\end{equation}                     
and for the central site Green's function
\begin{equation}
  g^R_c(\epsilon)=\frac{1}{\epsilon-\epsilon_c+i0}.
\end{equation}
In the presence of coupling between electrodes and the central site
the Dyson equation yields the central site Green's function in terms
of the self-energy
\begin{equation}
  G^R_c(\epsilon) = \frac{1}{\epsilon-\epsilon_c-\Sigma^R_c(\epsilon)}.
\end{equation}
The escape rate from the central site is given by the imaginary part
of the self energy
\begin{eqnarray}
  \Gamma(\epsilon)&=& -\Im m \Sigma^R_c(\epsilon) \nonumber \\
  &=& \pi \sum_{{\bf p}\sigma} |V_{{\bf p}\sigma}|^2
  \delta(\epsilon-\epsilon_{{\bf p}\sigma}).
\end{eqnarray}
If the hopping matrix elements vary slowly with energy in the
resonance region, $\epsilon \sim \epsilon_c$, and the energy bands in
the reservoirs are wide so that also the electrode density of states,
$N_{\sigma}(\epsilon)=\sum_{\bf p} \delta(\epsilon-\epsilon_{{\bf
    p}\sigma})$, has a weak energy dependence, we can neglect the
energy dependence of the escape rate, and by analyticity we have
\begin{equation}
  \Sigma^R_c(\epsilon) = \int_{-\infty}^{\infty}
  \frac{d\epsilon'}{\pi} \,
  \frac{\Gamma(\epsilon')}{\epsilon-\epsilon'+i0}, 
\end{equation}
which implies that the real part of the self energy vanishes in the
wide band limit.

The wide band approximation simplifies the problem considerably as the
time dependent Green's function becomes a decaying exponential
\begin{equation}
  \label{eq:widebandgreensf}
  G^R_c(t,t') = -i\theta(t-t')e^{-\frac{i}{\hbar}(\epsilon_c - i
    \Gamma)(t-t')},
\end{equation}
and the Green's function satisfies, for times $t>t''>t'$, the relation
\begin{equation}
  \label{eq:group}
  G^R_c(t,t'')\,G^R_c(t'',t')=G^R_c(t,t').
\end{equation}
This group property leads to a tremendous simplification of the
interacting problem as the tunneling dynamics at the central site and
the environment dynamics decouple. If the group property,
eq.(\ref{eq:group}), is valid we have for the central site amplitude
correlations
\begin{eqnarray}
  \label{eq:siteprob}
  \lefteqn{\langle  \hat{G}_c^A(t_4,t_3) \hat{G}_c^R(t_2,t_1)
  \rangle} \nonumber \\
  &=& G^A_c(t_4,t_3)G^R_c(t_2,t_1)
  Z(t_4,t_3,t_2,t_1),
\end{eqnarray}
where we have introduced the influence function describing the effect
of the environment
\begin{equation}
  \label{eq:influence_def}
  Z(t_4,t_3,t_2,t_1)=\langle (\tilde{T}e^{\frac{i}{\hbar}
    \int_{t_4}^{t_3}dt\ X(t)})
  (Te^{-\frac{i}{\hbar}\int_{t_1}^{t_2}dt\ X(t)})\rangle
\end{equation} 
with the environment variable in the interaction picture
\begin{equation}
  X(t)=e^{\frac{i}{\hbar}H_bt} X\ e^{-\frac{i}{\hbar}H_bt}.
\end{equation}
and $T$ and $\tilde{T}$ denotes the time and anti-time ordering
operators, respectively. The decoupling of the particle and
environment degrees of freedom is visualized using Feynman diagrams in
figure \ref{fig:decoupling}.

Because of the group property, eq.(\ref{eq:group}), the central site
dynamics has a special behavior: the lifetime of the central site
state is independent of the coupling to the environment, as the
probability, $P_c(t)$, for the particle to remain at the central site
after a time span $t$, if initially at the central site, is
\begin{equation}
  P_c(t) = \langle  \hat{G}_c^A(0,t) \hat{G}_c^R(t,0) \rangle,
\end{equation}
and noting that $T e^{-\frac{i}{\hbar} \int_0^t d\bar{t}\;
  X(\bar{t})}$ is the interaction picture time evolution operator for
the bath, and consequently unitary, we have $Z(0,t,t,0) = 1$.
Therefore by eq.(\ref{eq:siteprob}) we have for the staying
probability
\begin{equation}
  P_c(t) = \langle  G_c^A(0,t) G_c^R(t,0) \rangle = e^{-
    \frac{2\Gamma}{\hbar} t} ,
\end{equation}
which is independent of the coupling to the environment.\cite{Wilkins}

Before we go on to calculate the influence function $Z$, which
contains all information about the influence of the environment for
various environmental states, we briefly discuss the transmission
problem in the absence of coupling to the environment. We shall
explicitly assume the exponential form eq.(\ref{eq:widebandgreensf})
for the decay amplitude.

If we for the moment neglect the coupling to the environment we have
for the transition probability
\begin{eqnarray}
  \label{eq:nctransprob}
  \lefteqn{P_{{\bf p}',l \rightarrow {\bf p},r}(t)} \nonumber \\
  &=& \frac{1}{\hbar^4}|V_{{\bf p},r}|^{2}
  |V_{{\bf p}',l}|^{2}
  \int_0^t \!\! dt_1 \int_{t_1}^t \!\! dt_2 \int_0^t \!\! dt_3
  \int_0^{t_3} \!\!\! dt_4 \nonumber \\
  & & e^{\frac{i}{\hbar}
    \epsilon_{{\bf p},r}(t_2-t_3)
    +\frac{i}{\hbar}\epsilon_{{\bf p},l}
    (t_4-t_1)} \nonumber \\
  & &e^{-\frac{i}{\hbar}(\epsilon_c-i\Gamma)(t_2-t_1)
    +\frac{i}{\hbar}(\epsilon_c+i\Gamma)(t_3-t_4)}.
\end{eqnarray}
The integrations are readily done and we obtain for the transition
probability per unit time
\begin{eqnarray}
  \label{eq:nctransprobput}
  w_{{\bf p}',l\rightarrow {\bf
          p},r} &=& \lim_{t\rightarrow \infty}
  \frac{P_{{\bf p}',l\rightarrow
      {\bf p},r}(t)}{t} \nonumber \\
  &=& \frac{2\pi}{\hbar}
  \frac{| V_{{\bf p},r} |^2 |
    V_{{\bf p}',l}|^2 }
  {(\epsilon_{{\bf p}',l}
    -\epsilon_c)^2+\Gamma^2} \delta(\epsilon_{{\bf
          p},r} -\epsilon_{{\bf p}',l}),
\end{eqnarray}
or equivalently for the transmission coefficient
\begin{eqnarray}
  T(\epsilon,\epsilon ') &=& \frac{4 \Gamma_l
    \Gamma_r}{(\epsilon' - \epsilon_c)^{2} + \Gamma^{2}}\ 
  \delta(\epsilon-\epsilon') \nonumber \\
  &=& \frac{2\Gamma_l\Gamma_r}{\Gamma}A_c(\epsilon') \delta(\epsilon -
  \epsilon'),
\end{eqnarray}
where we have introduced the central site spectral weight function
\begin{eqnarray}
  A_c(\epsilon') &=& i\left(G^R_c(\epsilon') - G^A_c(\epsilon')\right)
  \nonumber \\
  &=& \frac{2\Gamma}{(\epsilon'-\epsilon_c)^2+\Gamma^2}
\end{eqnarray}
and the left and right escape rates
\begin{equation}
  \Gamma_{\sigma} = \pi \sum_{\bf p} |V_{{\bf p}\sigma}|^2
  \delta(\epsilon - \epsilon_{{\bf p}\sigma}).
\end{equation}
We observe that the transmission coefficient has the expected resonant
character of the Breit-Wigner formula\cite{BreitWigner}.
                                    
In the absence of coupling to the environment we therefore have the
contribution to the conductance at energy $\epsilon'$
\begin{equation}
  \label{eq:breit-wigner}
  G(\epsilon')= \frac{4e^{2}}{\pi \hbar}\frac{\Gamma_l
    \Gamma_r}{(\epsilon' - \epsilon_c)^{2} + \Gamma^{2}}  .
\end{equation}                

Having discussed briefly the uncoupled case, we now turn to calculate
the influence function Z for various environmental states. We shall
invoke the assumption of a wide band width allowing for the decoupling
expressed by eq.(\ref{eq:siteprob}), and obtain in this approximation
the following expression for the transmission coefficient
\begin{eqnarray}
  \label{eq:transprob2}
  T(\epsilon,\epsilon ') & = & \lim_{t\rightarrow \infty} \frac{1}{t}
  \; \frac{2\Gamma_l \Gamma_r}{\pi \hbar^3}
  \int_0^t \!\! dt_1 \int_{t_1}^t \!\! dt_2 \int_0^t \!\! dt_3
  \int_0^{t_3} \!\!\! dt_4 \nonumber \\
  && Z(t_4,t_3,t_2,t_1)
  e^{\frac{i}{\hbar} \epsilon(t_2-t_3) +\frac{i}{\hbar}\epsilon'
    (t_4-t_1)} \nonumber \\
  &&e^{-\frac{i}{\hbar}(\epsilon_c-i\Gamma)(t_2-t_1)
    +\frac{i}{\hbar}(\epsilon_c+i\Gamma)(t_3-t_4)}  .
\end{eqnarray}                                              
In the wide band limit we notice the integral of the transmission
coefficient is related to the staying probaility at the central site
\begin{equation}
  \int_{-\infty}^{\infty} \!\! d\epsilon'\, T(\epsilon') =
  \frac{8\pi\Gamma_l \Gamma_r}{\hbar} \int_{0}^{\infty} \!\! dt \,
  P_c(t),
\end{equation}
and we can therefore conclude that the integral of $T(\epsilon')$ is
unaffected by the coupling to the environment since the staying
probability, $P_c(t)$, is unaffected.

\section{Closed time path formulation for the influence function}
\label{sec:path}
                                             
In order to calculate the influence of the environment on the
otherwise coherent quantum process, the task is to calculate the
influence function given in eq.(\ref{eq:influence_def}). The first
step in achieving such a goal is to transform the operator expression
into an analytical one.  A general and convenient method for performing
this transformation has been devised by Schwinger \cite{Schwinger}.
The key point to note for employing this method for the present
purpose is to note that the influence function Z quite generally can
be expressed in terms of a closed time path generating functional
provided the "force" is chosen properly as
\begin{equation}
  \label{eq:genfunc}
  Z(t_4,t_3,t_2,t_1) = \langle T_C\,e^{- \frac{i}{\hbar}
    \int_c^{} d \tau \, f(\tau) X(\tau)}\rangle | _{f(\tau) =
    f^{0}(\tau)},
\end{equation}
where $C$ is the closed time path extending from $-\infty$ to
$+\infty$ along $C_1$ and back again along $C_2$, as illustrated in
figure \ref{fig:path}, and $T_C$ is the contour ordering symbol,
ordering the environment operators $X(\tau)$ according to their
contour label position on the contour $C$ (earliest to the right) and
$f^{0}(\tau)$ is the function on the upper and lower branches of the
contour specified by
\begin{equation}
  f^0(\tau) = \left\{
  \begin{array}{lcll}
    f^0_1(t)&=&\theta(t_2-t) - \theta(t_1-t), & t=\tau \in C_1 \\
    f^0_2(t)&=&\theta(t_3-t) - \theta(t_4-t), & t=\tau \in C_2
  \end{array}
  \right..
\end{equation}
The closed time path generating functional
\begin{equation}
  Z[f]\equiv
  \,\langle T_C\,e^{-\frac{i}{\hbar}\int_c{} \!
    d\tau\,f(\tau)\,X(\tau)}\rangle
\end{equation}
is therefore the quantity of interest.

The influence of the environment only appears through the state of the
environment in the generating functional, and different environments
can now be handled on an equal footing corresponding to just
substituting the proper environmental state, i.e., the corresponding
statistical operator.
     
In the present formulation we note that corresponding to the influence
of an environment on a given physical quantity there corresponds a
function, here denoted the proper "force", which when inserted into
the generating functional completely specifies the influence of the
environment.

In the following section we calculate the generating functional and
the influence function for various environmental states.

\section{The influence function for typical environmental states}
\label{sec:environments}

In the preceding section we showed that in the presented model the
influence functional refers only explicitly to the environmental
degrees of freedom and is characterized by a single "force". This
simplifying feature will allow us to obtain closed expressions for the
influence function for typical environmental states.

\subsection{Thermal environment}
                               
We first discuss the quite common physical situation where the
environment is acting as a heat bath. This could, for example, be the
case in question where the Hamiltonian is thought to represent a
resonant tunneling structure, the lattice degrees of freedom of the
crystal acting as the heat bath.

A heat bath is characterized by a single macroscopic parameter, its
temperature $T$, and the environmental state is in this case described
by the equilibrium statistical operator
\begin{equation}
  \rho_{b}=\rho_{T} = \frac{e^{-H_{b}/k_{B}T}}
  {tr e^{-H_{b}/k_{B}T}}.
\end{equation}
The average of the generating functional is then Gaussian, yielding
the quadratic form
\begin{equation}
  \label{eq:wicks}
  Z[f]=\exp\left\{ -\frac{i}{2\hbar^{2}}\int_c{}\!d\tau\int_c{}
  \!d\tau' \,f(\tau) D(\tau,\tau') f(\tau') \right\},
\end{equation}
where
\begin{equation}
  D(\tau,\tau')=-i\langle T_C(X(\tau)\,X(\tau'))\rangle
\end{equation}
is the contour ordered bath Green's function.

It is convenient for the physical interpretation to split the exponent
appearing in the generating functional into real and imaginary parts
\cite{albert}
\begin{eqnarray}
  \label{eq:riparts}
  Z[f]&=&\exp\bigg\{-\frac{i}{\hbar^{2}}\int_{-\infty}^{\infty}\!\!dt
  \int_{-\infty}^{\infty}\!\! dt' \,
  [2f_{-}(t)D^R(t,t')\, f_{+}(t') \nonumber \\
  &&\hspace{15mm} + f_{-}(t)D^{K}(t,t')\, f_{-}(t')] \bigg\},
\end{eqnarray}
where
\begin{equation}
  f_{\pm}(t)=\frac{1}{2}(f_1(t)\pm f_2(t)),
\end{equation}
and
\begin{equation}
  D^R(t,t')=-i\theta(t-t')\langle [X(t),X(t')]\rangle
\end{equation} 
is the retarded bath propagator, and
\begin{equation}
  D^{K}(t,t')=-i\langle \{X(t),X(t')\}\rangle
\end{equation}
is the correlation or Keldysh bath propagator \cite{RammerSmith}.

For the present thermal case we have for the retarded Green's function
\begin{equation}
  D^R(t,t')=-\frac{\hbar^{2}}{2}\theta(t-t')\int_0^{\infty}
  \!\!d\omega \, J(\omega) \sin\omega(t-t')  ,
\end{equation}
where
\begin{equation}
  J(\omega)=\frac{4}{\hbar^{2}}\sum_{\alpha}\,\lambda^{2}_{\alpha}
  \delta(\omega-\omega_{\alpha}) = - \frac{4}{\pi\hbar^2} \Im m
  D^R(\omega) 
\end{equation}
is the spectral function completely characterizing the influence of
the microscopic degrees of freedom of the bath.  For the correlation
function we similarly have
\begin{equation}
  D^{K}(t,t')=-i\frac{\hbar^2}{2}\int_0^{\infty}
  \!\!d\omega\,J(\omega)\, \coth\frac{\hbar\omega}{2k_{B}T}
  \cos\omega(t-t').
\end{equation}                       
The Fourier transforms of the retarded and the correlation propagator
is connected according to the fluctuation-dissipation theorem
\begin{equation}
  D^{K}(\omega)=2i\Im m D^R(\omega)\coth\frac{\hbar\omega}{2k_{B}T}.
\end{equation}

The calculation of the influence of the thermal reservoir on the
transition probability has now been reduced to the performance of
simple integrals which are readily done and we obtain for the
influence function
\begin{equation}
  Z(t_4,t_3,t_2,t_1)=Z^R(t_4,t_3,t_2,t_1)
  \,Z^{K}(t_4,t_3,t_2,t_1)  ,
\end{equation}
where the contribution from the retarded bath propagator is given by
\begin{equation}
  Z^R(t_4,t_3,t_2,t_1)=e^{\frac{i}{4}
    \lambda(t_2-t_1+t_4-t_3)}\,
  \tilde{Z}^R(t_4,t_3,t_2,t_1)  ,
\end{equation}
with
\begin{equation}
  \tilde{Z}^R(t_4,t_3,t_2,t_1) = \exp\left\{ \frac{i}{4}
  \int_0^{\infty}\!\!d\omega\, \frac{J(\omega)}
  {\omega^{2}}S_{\omega}(t_4,t_3,t_2,t_1)\right\}  ,
\end{equation}
specified by the function
\begin{eqnarray}
  S_{\omega}(t_4,t_3,t_2,t_1) & = & - \sin\omega (t_2 - t_1)
  - \sin\omega (t_3-t_2) \nonumber \\
  &&+ \sin\omega (t_4-t_2)  + \sin\omega (t_3-t_1) \nonumber \\ 
  &&- \sin\omega (t_4-t_1) + \sin \omega  (t_3-t_4) ,
\end{eqnarray}
and the effective coupling constant given by
\begin{equation}
  \lambda = \int_0^{\infty}\!\!d\omega \, \frac{J(\omega)}{\omega}
   .
\end{equation}
The contribution from the correlation bath propagator is similarly
given by
\begin{eqnarray}
  Z^{K}(t_4,t_3,t_2,t_1)&=&\exp\bigg\{-\frac{1}{4}
  \int_0^{\infty} \!\!d\omega \, \frac{J(\omega)}{\omega^{2}}
  \coth\frac{\hbar\omega}{2k_{B}T} \nonumber \\
  &&\hspace{20mm} C_{\omega}(t_4,t_3,t_2,t_1)\bigg\},
\end{eqnarray}
where
\begin{eqnarray}
  C_{\omega}(t_4,t_3,t_2,t_1) & = & 2-\cos\omega(t_1-t_2) -
  \cos\omega(t_2-t_3) \nonumber \\
  &&+ \cos\omega(t_1-t_3) + \cos\omega(t_2-t_4) \nonumber \\
  &&- \cos\omega(t_1-t_4) -  \cos\omega(t_4-t_3).
\end{eqnarray}
The two real functions $C_{\omega}$ and $S_{\omega}$ are the real and
imaginary parts of the complex function
\begin{eqnarray}
  f_{\omega}(t_4,t_3,t_2,t_1) &=& 2 - e^{i\omega(t_2-t_1)} -
  e^{i\omega(t_3-t_2)} \nonumber \\
  &&+ e^{i\omega(t_3-t_1)} + e^{i\omega(t_4-t_2)} \nonumber \\
  &&- e^{i\omega(t_4-t_1)} - e^{i\omega(t_4-t_3)}.
\end{eqnarray}

Since the environment is in thermal equilibrium the influence function
$Z(t_4,t_3,t_2,t_1)$ in fact depends only on three independent
variables
\begin{eqnarray}
  \tau   &=& t_2-t_1 \nonumber \\
  \tau'  &=& t_3-t_4 \nonumber \\
  \tau'' &=& t_3-t_2,
\end{eqnarray}
the above choice being determined by the original time labeling of the
Green's functions.

In terms of the independent variables we then obtain for the
transition probability per unit time
\begin{eqnarray}
  \lefteqn{w_{{\bf p}',l \rightarrow {\bf p},r} =
    \lim_{t\rightarrow\infty} \frac{P(t)}{t}} \nonumber \\
  & = &\frac{| V_{{\bf p},r} | ^{2} | V_{{\bf
        p}',l}|^{2}}{\hbar^{4}}
   \int_{-\infty}^{\infty} \! d\tau'' \int_0^{\infty} \!\!  d\tau'
  \int_0^{\infty} \!\! d\tau\, \tilde{Z} (\tau, \tau ', \tau '')
  \nonumber \\
  & & \exp\left\{-\frac{i}{\hbar}
  (\epsilon_{{\bf p},r}-\epsilon_{{\bf p}',l}) \tau '' -
  \frac{i}{\hbar} (\epsilon_{{\bf p}',l} -
  \tilde{\epsilon_c} -i \Gamma ) \tau ' \right. \nonumber \\
  &&\hspace{10mm}\left. + \frac{i}{\hbar}
  (\epsilon_{{\bf p}',l} - \tilde{\epsilon_c} +
  i \Gamma )\tau \right\},
\end{eqnarray}
where the influence function in terms of the three independent
variables again is split into the two distinct parts
\begin{equation}
  \tilde{Z}(\tau, \tau ', \tau '') =\tilde{Z}^R(\tau, \tau', \tau
  '')\, Z^{K}(\tau, \tau', \tau''),
\end{equation}                                    
where 
\begin{equation}
  \tilde{Z}^R(\tau, \tau ', \tau '') = \exp\left\{
  \frac{i}{4}\int_0^{\infty}
  \!\! d\omega\,\frac{J(\omega)}{\omega^{2}} S_{\omega}
  (\tau, \tau ', \tau '') \right\}  ,
\end{equation} 
with
\begin{eqnarray}
  S_{\omega}(\tau, \tau ', \tau '') & = & -\sin\omega\tau +
  \sin\omega\tau'  \nonumber \\
  &&- \sin\omega\tau''  +\sin\omega(\tau''-\tau') \nonumber \\
  &&+ \sin\omega(\tau''+\tau) - \sin\omega(\tau''+\tau-\tau'),
\end{eqnarray}
and the correlation part
\begin{eqnarray}
  Z^{K}(\tau, \tau ', \tau '')&=&
  \exp\left\{-\frac{1}{4}\int_0^{\infty}\!\!d\omega\,
  \frac{J(\omega)}{\omega^{2}} C_{\omega}(\tau, \tau ', \tau '')
  \right. \nonumber \\
  && \left. \hspace{25mm} \coth\frac{\hbar\omega}{2k_{B}T}\right\},
\end{eqnarray}
where
\begin{eqnarray}
  C_{\omega}(\tau, \tau ', \tau '') & = & 2 - \cos \omega \tau-\cos
  \omega \tau' \nonumber \\
  && - \cos \omega \tau''  + \cos\omega(\tau''-\tau' )\nonumber \\
  && + \cos \omega (\tau+\tau'') - \cos \omega (\tau''+\tau-\tau').
\end{eqnarray}
The central site energy is shifted downwards according to
\begin{equation}
  \tilde{\epsilon}_c=\epsilon_c-\frac{\hbar\lambda}{4}
\end{equation}
similar to the negative polaronic energy shift.

For the transmission coefficient in the case of a thermal environment
we then get
\begin{eqnarray}
  \label{eq:transcoeff}
  T(\epsilon,\epsilon') & =
  &\frac{2}{\pi \hbar^3}\Gamma_l\Gamma_r
  \int_{-\infty}^{\infty} \! d\tau '' \int_0^{\infty} \!\! d\tau '
  \int_0^{\infty} \!\! d \tau\, \tilde{Z} (\tau, \tau ', \tau '')
  \nonumber \\ & & \exp\left\{ - \frac{i}{\hbar} (\epsilon -\epsilon')
  \tau '' - \frac{i}{\hbar} (\epsilon' - \tilde{\epsilon_c} -i
  \Gamma ) \tau ' \right. \nonumber \\
  &&\left. \hspace{8mm} + \frac{i}{\hbar} (\epsilon' - \tilde{\epsilon_c}
  + i \Gamma )\tau \right\}.
\end{eqnarray}
This expression has been studied perturbatively in the case of the
Einstein model \cite{Wilkins}, and also in terms of elastic and
inelastic channels \cite{Glazman}. In the following we investigate the
total transmission coefficient in the thermal case and obtain explicit
expressions for various limiting situations.
 
\subsubsection{The total transmission coefficient}

If we are not interested in the energetics of the arriving particles
in the right electrode, but only in the number of arriving particles,
only the total transmission coefficient is relevant.  Such a situation
arises for instance in the case where we can neglect any effect of the
Pauli principle in the right electrode, corresponding to the situation
of a highly biased left electrode in which case the left-going current
is zero. In this situation the current of arriving particles is
\begin{equation}
  I = e \int_0^{\infty}\!\!d\epsilon' \, T(\epsilon ')
  f_L(\epsilon').
\end{equation}          

In such a case we only need the expression for the total transmission
coefficient which, according to
eq.(\ref{eq:transcoeff},\ref{eq:tot_trans_coeff}), is given by
\begin{eqnarray}
  \label{eq:tcoeff}
  T(\epsilon ') & = & \frac{4}{\hbar^2} \Gamma_l\Gamma_r
  \int_0^{\infty} \!\! d\tau ' \int_0^{\infty} \!\! d \tau\, 
  \tilde{Z}(\tau,\tau') \nonumber \\ & & \exp\left\{-\frac{i}{\hbar}
  (\epsilon ' - \tilde{\epsilon_c} -i \Gamma ) \tau ' +
  \frac{i}{\hbar} (\epsilon ' - \tilde{\epsilon_c} + i \Gamma )\tau
  \right\}, \nonumber \\
  &&
\end{eqnarray}
where
\begin{equation}
  \tilde{Z}(\tau,\tau') = \tilde{Z}^R(\tau,\tau') Z^K(\tau,\tau'),
\end{equation}
is specified by
\begin{eqnarray}
  \tilde{Z}^R(\tau,\tau')&=&\tilde{Z}^R(\tau,\tau',0) \nonumber\\
  &=& \exp\left\{
  -\frac{i}{4}\int_0^{\infty}
  \!\!d\omega\,\frac{J(\omega)}{\omega^{2}} 
  \sin\omega(\tau-\tau') \right\},
\end{eqnarray}
and the correlation part
\begin{eqnarray}
  Z^K(\tau,\tau') &=& Z^K(\tau,\tau',0) \nonumber \\
  &=& \exp\bigg\{
  - \frac{1}{4}\int_0^{\infty} \!\!
  d\omega\,\frac{J(\omega)}{\omega^{2}} \coth\frac{\hbar
    \omega}{2k_{B}T}  \nonumber \\
  &&\hspace{10mm} ( 1 - \cos\omega(\tau-\tau') ) \bigg\}.
\end{eqnarray}     
Since the environment is in the thermal equilibrium state, the influence
function for the total transmission coefficient only depends on one
time variable, and the expression for the total transmission
coefficient can be reduced to a single integral by introducing the
mean and relative time variables
\begin{eqnarray}
  t_m &=& \frac{\tau + \tau'}{2} \nonumber \\
  t &=& \tau - \tau'  .
\end{eqnarray}
For the integration region we observe
\begin{equation}
  \int_0^\infty d\tau \int_0^\infty d\tau' = \int_{-\infty}^{\infty} dt
  \int_{\frac{|t|}{2}}^\infty dt_m,
\end{equation}
and performing the integration over the mean time we finally obtain
\begin{eqnarray}
  \label{eq:th_env_trans}
  T(\epsilon ') &=& \frac{2}{
    \hbar}\frac{\Gamma_l\Gamma_r}{\Gamma} \int_0^{\infty}
  \!\! dt\, e^{-\frac{\Gamma t}{\hbar}} Z^K(t) \nonumber \\
  &&\left[ \tilde{Z}^R(t)\exp\left\{\frac{i}{\hbar} (\epsilon' -
  \tilde{\epsilon_c})t \right\} + c.c. \right],
\end{eqnarray}
where
\begin{equation}
  \label{eq:retarded_influence}
  \tilde{Z}^R(t)= \exp\left\{ -\frac{i}{4}\int_0^{\infty} \!\!
  d\omega\, \frac{J(\omega)}{\omega^{2}} \sin\omega t \right\},
\end{equation}
and
\begin{equation}
  \label{eq:keldysh_influence}
  Z^K(t)= \exp\left\{ - \frac{1}{4}\int_0^{\infty} \!\!
  d\omega\, \frac{J(\omega)}{\omega^{2}}\coth\frac{\hbar \omega}
  {2k_{B}T}\ ( 1 - \cos\omega t ) \right\}.
\end{equation}                  

In the case where the oscillators all have the same frequency, the
Einstein model, the spectral function takes the form
\begin{equation}
  J(\omega) = \frac{4 \lambda_0^{2} }{\hbar^{2} } \delta(\omega -
  \omega_0 )
  \label{eq:einstein}
\end{equation}
relevant, e.g., to optical phonons. In this case, we deduce from
eqs.(\ref{eq:th_env_trans}-\ref{eq:keldysh_influence}) the total
transmission coefficient
\begin{eqnarray}
  T(\epsilon ') &=& \frac{2\Gamma_l\Gamma_r}{\hbar\Gamma}
  \int_0^{\infty} \!\! dt\, \exp\left\{ - \left( \frac{ \lambda_0
    }{\hbar\omega_0 } \right)^2 ( 1 + 2 n(\omega_0 ) ) \right\}
  \nonumber \\
  &&\Bigg[\exp\left\{ \left( \frac{ \lambda_0 }{\hbar\omega_0 } \right)^2 ( 1
  + n(\omega_0) ) e^{- i \omega_0 t } \right\} \nonumber \\
  &&\exp\left\{ \left(
  \frac{ \lambda_0 }{\hbar\omega_0 } \right)^2 n(\omega_0 ) e^{i
    \omega_0 t } \right\} \nonumber \\
  && \exp\left\{ - \frac{1}{\hbar} ( \Gamma - i
  (\epsilon' - \tilde{\epsilon_c}) )t \right\} + c.c. \Bigg],
\end{eqnarray}
where $n(\omega)$ is the Bose function
\begin{equation}
  n(\omega) = \frac{1}{ e^{\hbar\omega/k_{B}T} - 1 }.
\end{equation}
Expanding the exponential functions containing $e^{ i \omega_0 t }$
and $ e^{ - i \omega_0 t }$, and performing the integration over $t$
we obtain
\begin{equation}
  T(\epsilon ') = 4 \Gamma_l\Gamma_r
  \sum_{n=-\infty}^{\infty} \frac{P_n(T) } {
    (\epsilon'-\tilde{\epsilon_c}-n\hbar\omega_0)^{2}+\Gamma^{2} }
   ,
\end{equation}
where the temperature dependence is specified by
\begin{eqnarray}
  \label{eq:lineweight}
  P_n(T) & = & \exp\left\{ - \left(\frac{ \lambda_0
    }{\hbar\omega_0 }\right)^2 [ 1 + 2 n(\omega_0 ) ] \right\}
  \nonumber \\ 
  & &\sum_{n_1,n_2=0}^{\infty} \raisebox{1.5ex}{\hspace{-0.7em}$\prime$}
  \hspace{0.7em} \frac{1}{n_1! n_2 ! } \left[
  \left(\frac{\lambda_0 }{\hbar\omega_0 } \right)^{2} [ 1 +
    n(\omega_0 )] \right]^{n_1} \nonumber \\
  & & \hspace{6em} \left[ \left(\frac{\lambda_0
      }{\hbar\omega_0 } \right)^{2} n(\omega_0 ) \right]^{n_2},
\end{eqnarray}
the prime restricting the summation to the terms for which
$n_1-n_2=n$. Transmission can take place with absorption or emission
of oscillator quanta giving rise to additional resonance peaks, the
function $P_n(T)$ determining the relative weight of the peaks.  The
energy dependence of the transmission coefficient is illustrated in
figure \ref{fig:einstein}. In particular, in the low temperature
limit, $k_BT < \hbar\omega_0$, where the Bose function vanishes,
$n(\omega) \rightarrow 0$, we can ignore all terms containing powers
of $n(\omega)$ in eq.(\ref{eq:lineweight}), and obtain at zero
temperature
\begin{equation}
  \label{eq:lineweight_T=0}
  P_n(0) = \left\{ 
  \begin{array}{ll}
    \exp\left\{ - \left( \frac{ \lambda_0 }{\hbar\omega_0
    }\right)^2 \right\} \frac{1}{n!}
    \left(\frac{\lambda_0 }{\hbar\omega_0 }
    \right)^{2n}& n \geq 0 \\
    0 & n<0
  \end{array}
  \right.,
\end{equation}
reflecting the possibility of an electron off resonance to tunnel from
the left to the right reservoir via the central site by spontaneously
emitting a phonon, and the impossibility of gaining energy from a
zero temperature environment. We note that $P_n(0)$ is a Poisson
distribution characterized by its mean value
\begin{equation}
  \langle n \rangle = \sum_{n=0}^{\infty} n P_n(0)
  = \left( \frac{\lambda_0}{\hbar\omega_0} \right)^2.
\end{equation}

\subsubsection{Approximations conserving the integrated
  transmission probability}

To illustrate the general features of the systematic and fluctuation
influences of the environment we choose the spectral function in the
further calculations to have the form
\begin{equation}
  J(\omega) = \eta \omega \left( \frac{\omega}{\omega_c}
  \right)^{s-1} \exp\left(-\frac{\omega}{\omega_c } \right).
\end{equation}
We note, that $\eta$ is a dimensionless constant describing the
coupling strength between the central level and the environment, and
$\omega_c$ is the cut-off frequency for the oscillators.

There are several limits in which simple expressions for
the transmission coefficient may be worked out. These limits are the
broad resonance limit, $\Gamma>\hbar\omega_c$, the strong coupling
limit, $\eta > 1$, the high temperature limit, $k_{B}T >
(1+1/\eta)\hbar\omega_c$, and the weak coupling and low temperature
limit, $\eta < 1, \, k_BT < \frac{\hbar\omega_c}{\eta}$.

\paragraph{Broad resonance limit.}

In the broad resonance limit, $\Gamma > \hbar\omega_c$, a short time
expansion of the influence function is sufficient. The correlation
part of the influence function is therefore given by
\begin{equation}
  \label{eq:keldysh_broad}
  Z^K(t) = \exp\left\{-\frac{\kappa(T)}{8} t^{2} \right\},
\end{equation}
which controls the other part to become
\begin{equation}
  \label{eq:retarded_broad}
  \tilde{Z}^R(t) = \exp\left\{ -i \frac{\lambda}{4} t \right\} .
\end{equation}
Here $\kappa$ is defined as
\begin{equation}
  \kappa(T) = \int_0^{\infty} \!\! d\omega \, J(\omega)
  \coth\frac{\hbar\omega}{ 2k_{B}T }  .
\end{equation}
The total transmission coefficient, eq.(\ref{eq:th_env_trans}), is
therefore in the broad resonance limit given by
\begin{eqnarray}
  \label{eq:transbroad}
  T(\epsilon') &=& \frac{2\Gamma_l\Gamma_r
    }{\hbar \Gamma} \int_0^{\infty} \!\!\!\! dt \, \left[ e^{ -
  \frac{\kappa(T)}{8} t^{2} - \frac{1}{\hbar} [ \Gamma - i (\epsilon' -
  \epsilon_c) ] t} 
   + {\rm c.c.} \right].
\end{eqnarray}
We note that the polaron shift has canceled out, reflecting that the
escape rate out of the central site is so fast that any real part
environmental self-energy dressing effect is absent.

The integral in eq.(\ref{eq:transbroad}) can now be performed and we
obtain
\begin{eqnarray}
  \label{eq:broadtrans}
  T(\epsilon') &=& \sqrt{ \frac{2\pi}{\kappa(T)} }
  \frac{2\Gamma_l\Gamma_r }{\Gamma \hbar} \Bigg\{ e^{
    \frac{2}{\kappa(T) \hbar^{2} } (\Gamma - i(\epsilon' -\epsilon_c)
    )^{2} } \nonumber \\
  && \left[ 1- \Phi\left(\sqrt{\frac{2}{\kappa(T)\hbar^{2} } } [
  \Gamma - i(\epsilon' -\epsilon_c) ] \right)\right] + {\rm c.c.} \Bigg\},
\end{eqnarray}
where $\Phi(z)$ is the probability integral, the error function.  We
note, that in the above case the resonant line shape is no longer of
the Lorentz type. The energy dependence in the broad resonance limit
of the transmission coefficient in the ohmic case, $s=1$, is
illustrated in figure \ref{fig:broad}.

\paragraph{Strong coupling limit.}

In the strong coupling limit, $\eta >1$, we only need to consider the
short time limit, $\omega_c t<1$, because at larger times the
influence function is exponentially small due to the short range of
the correlation part, $Z^K(t)$. Therefore, the correlation part of the
influence function is for the present case the same as in
eq.(\ref{eq:keldysh_broad}), and similarly the retarded part of the
influence function is specified by eq.(\ref{eq:retarded_broad}). The
expression for the total transmission coefficient is therefore the
same as the one given in eq.(\ref{eq:broadtrans}). However, the
validity condition here is the strong coupling criteria, $\eta>1$, and
there is no requirement on the escape rate $\Gamma$. We note again,
that in the short time limit approximation the retarded part of the
influence function has no effect, and the correlation part of the
influence function is for the present case the relevant one. In figure
\ref{fig:strong_coupling} the strong coupling approximation for the
transmission coefficient is plotted versus energy for the case of an
ohmic bath. We get in the strong coupling approximation for typical
parameter values, $\eta=10$, $\hbar\omega_c=\Gamma$ and
$2k_BT=\Gamma$, a $5\%$ too low transmission maximum which is
displaced to a slightly higher energy as compared to the exact result.

\paragraph{High temperature limit.}

When the temperature is high enough, $k_{B}T > (1 + 1/\eta)
\hbar\omega_c $, we can again use a short time approximation 
\begin{equation}
  Z^K(t) = \exp\left\{-\frac{\kappa'(T)}{8} t^2 \right\}
\end{equation}
where now
\begin{equation}
  \kappa'(T) = \frac{2k_{B}T}{\hbar} \int_0^{\infty} \!\! d\omega
  \, \frac{J(\omega) }{\omega},
\end{equation}
which again controls the retarded part of the influence function to be
\begin{equation}
  \tilde{Z}^R(t) = \exp\left\{-i \frac{\lambda}{4} t \right\}.
\end{equation}
Similar to the two previous cases, the expression for the total
transmission coefficient is
\begin{eqnarray}
  T(\epsilon') &=& \sqrt{ \frac{2\pi}{\kappa'(T)} }
  \frac{2\Gamma_l\Gamma_r }{\Gamma \hbar} \Bigg\{ e^{
    \frac{2}{\kappa'(T) \hbar^{2} } (\Gamma - i(\epsilon' -\epsilon_c)
    )^{2} } \nonumber \\
  &&\left[ 1- \Phi\left(\sqrt{\frac{2}{\kappa'(T)\hbar^{2} } } [
  \Gamma - i(\epsilon' -\epsilon_c) ] \right)\right] + {\rm c.c.} \Bigg\},
\end{eqnarray}
which has the same form as eq.(\ref{eq:broadtrans}), except that
$\kappa$ is replaced by $\kappa'$. The high temperature approximation
is compared to the exact result in figure \ref{fig:high_temp}, for the
parameter values $\eta=1$, $\hbar\omega_c=\Gamma$ and $2k_BT=10
\Gamma$. We note that the high temperature approximation gives a
slightly lower transmission maximum than the exact calculation yields.

\paragraph{Weak coupling and low temperature limit.}

When the coupling is weak, $\eta < 1$, and the temperature is low,
$k_{B}T < \hbar\omega_c/\eta$, the situation is different from the
previous cases. The weak coupling condition, $\eta<1$, forces the part
of the influence function, $\tilde{Z}^R$, to equal unity at all times,
$\tilde{Z}^R(t) \sim 1$. We also note that there is no contribution
from the correlation part in the short time limit due to the weak
coupling condition.  We need therefore to consider the long time
behavior of the correlation part. The reason for the condition $k_{B}T
< \hbar\omega_c/\eta$ is to avoid the high temperature regime where
only short times give a contribution.

The argument of the exponential in the expression for the correlation
part of the influnence function
\begin{equation}
  I_s(t) \equiv \int_0^{\infty} \!\! d\omega \, \frac{J(\omega)
    }{\omega^{2} } \coth\frac{\hbar\omega}{2k_{B}T} (1-\cos\omega t)
\end{equation}
approaches in the long time limit, $t > \max(\frac{1}{\omega_c},
\frac{\hbar}{2k_BT})$, the expression
\begin{equation}
  I_s(t) = \eta \frac{2k_{B}T}{\hbar} t (\omega_c t )^{1-s} \int_0^{\infty}
  \!\!\! dx \, x^{s-3} (1-\cos x),
\end{equation}
for the exponent region $0< s <2$. If $s> 2$, then $I_s(t)$ approaches
a constant in the long time limit, and there is therefore no
contribution from the correlation part in this case.  To be more
specific, we perform the integration for the ohmic case, $s=1$. In the
long time limit, $I_1(t)= \eta \frac{\pi k_BT}{\hbar}t$, and the ohmic
correlation part becomes
\begin{equation}
  Z^K(t) = \exp\left\{-\frac{\pi\eta k_{B}T}{4
    \hbar}t\right\}.
\end{equation}
Therefore, the total transmission coefficient is
\begin{eqnarray}
  T(\epsilon') &=& \frac{2 \Gamma_l\Gamma_r
    }{\hbar\Gamma} \int_0^{\infty} dt \Bigg[ \exp\bigg\{-\frac{1}{4}
  \eta\frac{\pi k_{B}T}{\hbar} t  \nonumber \\
  &&\hspace{15mm} - \frac{1}{\hbar} [ \Gamma - i (
  \epsilon' - \tilde{\epsilon_c} ) ] t \bigg\} + c.c. \Bigg].
\end{eqnarray}
The polaronic shift is no longer canceled as $\tilde{Z}^R \sim 1$.
Performing the integration we find for the total transmission
coefficient
\begin{equation}
  T(\epsilon') = \frac{\Gamma_{\rm eff}}{\Gamma}
  \frac{4 \Gamma_l\Gamma_r } { (\epsilon' - \tilde{\epsilon_c}
    )^{2} + \Gamma_{\rm eff}^{2} }  ,
\end{equation}
where
\begin{equation}
  \Gamma_{\rm eff} = \Gamma + \frac{\pi \eta k_B T}{4}.
\end{equation}
We note, that the Lorentzian character of the resonant transmission
coefficient is preserved in this limit, but the width and the height
of the resonance peak are different compared to the case where the
bath is absent.  Clearly, at high enough temperatures, $T >
\frac{4\Gamma}{\pi\eta k_B}$, the resonance peak will be strongly
reduced.  Therefore, although the coupling is weak, the result is
highly non-perturbative.  The weak coupling approximation in the
temperature range $\frac{4\Gamma}{\pi\eta} < k_BT <
\frac{\hbar\omega_c}{\eta}$ is compared to the exact result in figure
\ref{fig:wcltl2}, for parameter values $\eta=0.1$, $\hbar\omega_c=100
\Gamma$ and $2k_BT=100 \Gamma$.  For this parameter choice the
resonance peak is broadened approximatively five times compared to the
case where the environment is absent. Even though in the case
considered the coupling constant is not that small and the temperature
is not excessively low, the deviation of the approximately calculated
transmission maximum compared to the exact value is no more than
$10\%$.

We notice that the zeroth moment of the transmission coefficient,
\begin{equation}
  \int_{-\infty}^{\infty} \!\! d\epsilon' \, T(\epsilon') =
  \frac{4\pi\Gamma_l\Gamma_r}{\Gamma},
\end{equation}
is preserved exactly in all the approximations considered above. We
also notice that the first moment of the transmission coefficient,
\begin{equation}
  \frac{\Gamma}{4\pi\Gamma_l\Gamma_r} \int_{-\infty}^{\infty} \!\!
  d\epsilon' \, \epsilon' T(\epsilon') = \epsilon_c
\end{equation}
is preserved exactly by all the approximations except for the weak
coupling and low temperature approximation where the transmission
coefficient is symmetric around $\tilde{\epsilon}_c$, which is anyhow
close to $\epsilon_c$ because the coupling is weak.

\subsection{The fluctuating level model}

When discussing specific properties of a physical system, it is often
possible to neglect the quantum nature of the environment and
represent the effect of the environment by a collective classical
variable, which is appropriate when quantum fluctuations can be
neglected. A well known example is the excellent account of the atomic
spectra obtained by disregarding the quantum fluctuations of the
electromagnetic environment, except for cases where degeneracies are
only lifted by radiative corrections.  A counter example where the
effect of the environment is of pure quantum nature is of
course as easily recalled, that of stimulated emission.

In the case where the degree of freedom $X$ represents environmental
degrees of freedom collectively in the form of an external classical
potential, we can still use the presented method for calculating the
influence.  In this case the quantity $X$ just corresponds to a
classical potential, and is not an operator but just a c number.  We shall
in this section discuss the case where the potential
$X$ is a fluctuating quantity, the fluctuating level model.

We assume further that the fluctuations are Gaussian, and therefore
characterized by the lowest order correlations
\begin{eqnarray}
  \langle X(t)\rangle&=&c \nonumber \\
  \langle [X(t)- \langle X(t)\rangle][X(t')- \langle X(t')\rangle]\rangle
  &=& K(t-t').
\end{eqnarray}                            
The influence of the now classical environmental degree of freedom $X$
is given by the generating functional expression,
eq.(\ref{eq:genfunc}), however now the brackets simply denotes the
Gaussian average with respect to the fluctuating quantity $X$. We
therefore have for the generating functional, which in this case is
just the probability theory contour characteristic functional,
\begin{eqnarray}
  Z[f]&=&\exp\left\{ -
  \frac{i}{\hbar}\int_c  d\tau\,f(\tau)\langle X(\tau)\rangle
  \right. \nonumber \\
  &&\left. \hspace{8mm} -\frac{1}{2\hbar^{2}}\int_c  d\tau \int_c
  d\tau' \, 
  f(\tau)K(\tau-\tau')f(\tau') \right\}.
\end{eqnarray}
The reason we in the previous section made the distinction in the
effects of the environment as expressed in the split
\begin{equation}
  Z=Z^RZ^{K},
\end{equation}
where we distinguish between the retarded and correlation
contributions, is that they represent two distinct influences of the
environment.  The term $Z^R$ represents the systematic friction type
and energy renormalization influence and $Z^{K}$ the fluctuating part,
including thermal as well as quantum fluctuations.  For the thermal
environment discussed in the previous subsection, the two types of
influence were not independent, but related through the
fluctuation-dissipation theorem.

We now compare the thermal quantum environment with the classical
stochastic environment introduced above.  It follows directly in the
presented real-time approach that the fluctuating level model is in
one-to-one correspondence with the fluctuational aspect of the thermal
case with the prescription for the correlation part
\begin{equation}
  D^{K}(t,t') \mapsto -2iK(t-t'),
\end{equation} 
and substitution of one for $\tilde{Z}^R$, $\tilde{Z}^R\mapsto 1$. In
addition, we see that the polaronic shift corresponds to the average
displacement of the environment
\begin{equation}
  \frac{\hbar\lambda}{4} \mapsto -c,
\end{equation}
obtained from eq.(\ref{eq:tcoeff}).

The transmission coefficient for the stochastic environment is
obtained from
\begin{eqnarray}
  T(\epsilon') &=& \frac{4\Gamma_r\Gamma_l}{\hbar^2}
  \int^{\infty}_0 \!\!
  d\tau' \int^{\infty}_0 \!\! d\tau\, Z^{K}(\tau,\tau') \nonumber \\
  &&\exp\left\{\frac{i}{\hbar} (\epsilon' -
  \tilde{\epsilon}_c + i\Gamma ) \tau 
  -\frac{i}{\hbar} (\epsilon' - \tilde{\epsilon}_c - i\Gamma ) \tau'
  \right\}
\end{eqnarray}
where
\begin{equation}
  Z^{K}(\tau, \tau ') = \exp\left\{ -\frac{2}{\hbar^{2} }
  \int_{-\infty}^{\infty} dt \int_{-\infty}^{\infty} dt' f^0_{-}(t)
  K(t-t') f^0_{-}(t') \right\},
\end{equation}
and
\begin{eqnarray}
  f^0_{-}(t) &=& \frac{1}{2} [ (\theta(\tau - t ) - \theta(- t ) )
  \nonumber \\
  &&-(\theta( \tau - t ) - \theta(- \tau' + \tau - t )
  )]
\end{eqnarray}
and the correlation part $Z^{K}$ now depends on the explicit form of
the correlator $K$. For example, the high temperature form of the
correlation function $D^{K}$ for a thermal ohmic environment,
$J(\omega)=\eta \omega$, corresponds to a delta-correlator
\begin{equation}
  K(t-t')=\frac{\pi\hbar\eta k_{B}T}{2} \delta (t-t'),
\end{equation}
the white noise case. The fluctuating level model thus only
captures the correlation part $Z^{K}$ of the thermal environment and
neglects the systematic retarded part, except for its
average influence.
                                
For the fluctuating level model we therefore get for the influence
function in the Ohmic case
\begin{equation}
  Z^{K}(\tau,\tau')=\exp\left\{ -\frac{1}{4\hbar}\pi \eta k_{B}T |
  \tau-\tau' | \right\}  ,
\end{equation}
and for the transmission coefficient
\begin{equation}
  T(\epsilon') = \frac{\Gamma_{\rm eff}
    }{\Gamma } \frac{4\Gamma_l\Gamma_r }{ (\epsilon' -
    \tilde{\epsilon_c} )^{2} + \Gamma_{\rm eff}^{2} }.
\end{equation}
Comparing the result to the uncoupled case, we notice, besides the
average energy shift and a reduction of the peak, \cite{StoneLee} that
the resonance width $\Gamma$ of eq.(\ref{eq:breit-wigner}) is
broadened according to
\begin{equation}
  \Gamma\rightarrow\Gamma_{\rm eff}=\Gamma+\frac{\pi}{4 }\eta k_{B}T.
\end{equation}
We note that the transmission coefficient in this case is the same as
the one obtained for the thermal case in the weak coupling and low
temperature limit.

\subsection{The external field case}

Just as in the previous section we shall here consider the case where
the environment can be described classically. In addition we shall
assume, that in contrast to the previous section we know not only the
probability distribution of the potential, but in fact the actual
potential. In the following we shall therefore investigate the
situation where we are able to couple the electron to an external
field at the central site. In the present case of linear coupling it
is sufficient to consider the case where the central site energy level
changes harmonically in time.  The external potential is therefore
given by
\begin{equation}
  \label{eq:externalfield}
  X(t)=X_0\cos\omega_0t.
\end{equation}
Such a situation could for instance be realized in the case where the
Hamiltonian represents a small metallic grain with a single active
level whose energy can be changed by an external electric potential, and the
grain being coupled to large metallic electrodes.

In the case where the influence of the environment is represented by
the external potential given by eq.(\ref{eq:externalfield}) we obtain
for the influence function
\begin{eqnarray}
  Z(t_4, t_3, t_2, t_1) & = &
  \langle T_Ce^{-\frac{i}{\hbar}\int_cd\tau\,f(\tau)\,X(\tau)}\rangle
  |_{f=f^0} \nonumber \\
  &=& e^{-\frac{i}{\hbar}X_0\int_{-\infty}^{\infty}dt
    \,(f_1(t)-f_2(t))\cos\omega_0t} \nonumber \\
  & = & \exp \bigg\{ \frac{iX_0}{\hbar\omega_0} (\sin\omega_0
    t_1-\sin\omega_0t_2 \nonumber \\
  &&+\sin\omega_0t_3-\sin\omega_0t_4)\bigg\}.
\end{eqnarray}
This corresponds to the fluctuating level model for the case where the
potential $X(\tau)$ is known with certainty to be given by the expression
in eq.(\ref{eq:externalfield}).

For the transition probability per unit time we then obtain the
expression
\begin{eqnarray}
  \label{eq:exttransprob}
  w_{{\bf p}',l\rightarrow {\bf
          p},r} & = & \frac{| V_{{\bf p},r} | ^{2}
        | V_{{\bf p}',l} | ^2}{\hbar^{4}}  \int_{-\infty}^{\infty}
  \! d\tau '' \int_0^{\infty} \!\! d\tau ' \int_0^{\infty} \!\! d\tau 
  \nonumber \\
  & & \exp\left\{-\frac{i}{\hbar}
  (\epsilon_{{\bf p},r}-\epsilon_{{\bf p}',l}) \tau''
  \right. \nonumber \\
  &&\left. - \frac{i}{\hbar}(\epsilon_{{\bf p}',l} - \epsilon_c -i \Gamma
  ) \tau ' + \frac{i}{\hbar} (\epsilon_{{{\mbox{\scriptsize\bf
          p}}'},l} - \epsilon_c + i \Gamma )\tau \right\} \nonumber \\
  & & \lim_{t\rightarrow \infty}\frac{1}{t}\int_0^tdt_1 \exp
  \left\{ \frac{i}{\hbar}\frac{X_0}{\omega_0}
    (\Re e f^{\omega_0}_{\tau ,\tau' ,\tau ''} \sin\omega_0 t_1
  \right. \nonumber \\
  &&\left. +\Im m f^{\omega_0}_{\tau ,\tau' ,\tau ''} \cos\omega_0
    t_1)\right\},
\end{eqnarray}
where we have introduced the function
\begin{equation}
  \label{eq:timefunc}
  f^{\omega_0}_{\tau ,\tau' ,\tau ''} = 1 - e^{i\omega_0 \tau } +
  e^{i\omega_0(\tau + \tau'')} - e^{i\omega_0(\tau+\tau''
    -\tau')}.
\end{equation}

We now calculate the integral over $t_1$ in eq.(\ref{eq:exttransprob})
by expanding the exponential before performing the integration, and
therefore assume that $\omega_0$ is non-zero. The case of zero
$\omega_0$ corresponds to a constant external potential, $X(t)=X_0$,
which only yields a trivial shift of the resonant energy. The
transition probability per unit time can now be expressed as an
integral involving the zeroth order Bessel function,
$J_0$.\cite{Gradstheyn}
\begin{eqnarray}
  T(\epsilon,\epsilon ') & = & \frac{2\Gamma_l \Gamma_r}{\pi \hbar^3} 
   \int_{-\infty}^{\infty} \! d\tau '' \int_0^{\infty} \!\! d\tau
  ' \int_0^{\infty} \!\! d\tau \nonumber \\
  &&J_0\left( \frac{X_0}{\hbar \omega_0}|
  f^{\omega_0}_{\tau ,\tau' ,\tau ''} | \right) \nonumber \\
  & & \exp\left\{ -\frac{i}{\hbar} (\epsilon -\epsilon ')
  \tau '' - \frac{i}{\hbar} (\epsilon ' - \epsilon_c -i \Gamma )
  \tau' \right. \nonumber \\
  &&\left. + \frac{i}{\hbar} (\epsilon ' - \epsilon_c + i \Gamma )\tau
  \right\}.
\end{eqnarray}
If we had assumed that the external potential was a sine function,
$X(t)=X_0\sin \omega_0t$, we would have obtained the same transmission
coefficient for non-zero $\omega_0$, but we would of course not have
obtained a shift of the resonant energy in the case of $\omega_0=0$.
The total transmission coefficient is
\begin{eqnarray}
  T(\epsilon') & = & \frac{4\Gamma_l\Gamma_r}{\hbar^2} 
  \int_0^{\infty} \!\! d\tau \int_0^{\infty} \!\! d \tau' \, 
  J_0\left( \frac{X_0}{\hbar\omega_0}| f^{0}_{\tau ,\tau' ,\tau
    ''=0} | \right)\nonumber \\ 
  & & \exp\left\{ - \frac{i}{\hbar} (\epsilon ' -
  \epsilon_c -i \Gamma ) \tau ' + \frac{i}{\hbar} (\epsilon ' -
  \epsilon_c + i \Gamma )\tau \right\}. \nonumber \\
  &&
\end{eqnarray}
Noting that $| f^{\omega_0}_{\tau ,\tau' ,\tau ''=0} |^{2} = 4
\sin^{2}\left( \frac{\omega_0 (\tau-\tau') }{2} \right) $, and using
the summation formula for the Bessel function $J_0$\cite{Gradstheyn}
we obtain the total transmission coefficient\cite{Wingreen}
\begin{equation}
  \label{eq:exttottrans}
  T(\epsilon ') = 4 \Gamma_l \Gamma_r
  \sum_{n=-\infty}^{\infty} \frac{J_{n}^{2}(\frac{X_0}{\hbar
      \omega_0})} {(\epsilon ' - \epsilon_c - n\hbar
    \omega_0)^{2} + \Gamma^{2}}  .
\end{equation}
The total transmission coefficient has Lorentzian side bands at all
harmonics of $\hbar\omega_0$ with a relative weight determined by the
Bessel functions, and maximal peaks in the spectrum at $n \sim \pm
X_0/\hbar\omega_0 $. The transmission coefficient is plotted for the
parameters $\Gamma=0.2\hbar\omega_0$ and $X_0=4\hbar\omega_0$ in
figure \ref{fig:external}. For this parameter choice the first
sideband is almost missing because $\frac{X_0}{\hbar \omega_0}$ is
close to the first zero of the first order Bessel function.

\subsection{Coherent state environment}

In the following we shall investigate the model for the environment in
the coherent state. Such a model may describe transport through a
quantum dot coupled to an environment of coherent
phonons.\cite{Stanton}

The coherent state has the following representation in terms of the
vacuum state, $|0\rangle$, of the environment
\begin{equation}
  | \phi \rangle=N_{\phi}^{-\frac{1}{2}} \exp \{ \phi b^{\dag}_0\}|
  0\rangle,
\end{equation}
where
\begin{equation}
  N_{\phi}=\exp\{| \phi|^{2}\}
\end{equation}
is the normalization factor.

The state of the environment is then a pure state and the statistical
operator for the environment reduces to the projection operator
\begin{equation}
  \rho_{b}=P_{\phi}=\, | \phi\rangle\langle \phi |  .
\end{equation}

To calculate the influence of the environment in the case of a coherent
state we therefore need to evaluate the following expression for the
influence function
\begin{eqnarray}
  Z(t_4,t_3,t_2,t_1) &=& Z[f^{0}] \nonumber \\
  &=& N_{\phi}^{-1} \langle 0| \exp \{\phi^* b_0 \}\nonumber \\
  &&\left(T_C
  e^{-\frac{i}{\hbar}\int_c{}d\tau\,f^{0}(\tau)\,X(\tau)} \right)
  \exp \{\phi b^{\dag}_0 \}| 0\rangle. \nonumber \\
  &&
\end{eqnarray}                           

The bracket in the definition of the generating functional corresponds
therefore in this case to taking the expectation value with respect to
the coherent state
\begin{eqnarray}
  \label{eq:cohgenfunc}
  Z[f] & = &
  \langle T_Ce^{-\frac{i}{\hbar}\int_c{}d\tau\,f(\tau)\,X(\tau)}\rangle\, =
  \langle \phi| T_Ce^{-\frac{i}{\hbar}\int_cd\tau\,f(\tau)\,X(\tau)}
  | \phi\rangle \nonumber \\ & = & {N_{\phi}^{-1}}\langle 0 |
  e^{\phi^{\ast}b_0 } \; T_C e^{
    -\frac{i}{\hbar}\int_c{}d\tau\,f(\tau)\,X(\tau) } \; e^{\phi
    b^{\dag}_0 }| 0\rangle.
\end{eqnarray}                                                      

The matrix element appearing in eq.(\ref{eq:cohgenfunc}) is most
easily calculated by introducing the single mode generating functional
\begin{equation}
  \label{eq:singlemode}
  Z[f,f^*] = \langle T_Ce^{-\frac{i}{\hbar}\int_cd\tau
    \lambda_0 (f^*(\tau) \,
    b_0(\tau)+f(\tau)\,b^{\dag}_0(\tau))}\rangle  .
\end{equation}
We then notice, that we can rewrite the generating functional of
interest as
\begin{eqnarray}
  Z[f] &=& \langle \phi | T_C\,e^{-\frac{i}{\hbar}\int_cd\tau\,f(\tau)
    \, X(\tau)}| \phi\rangle \nonumber \\
  &=& Z[\tilde{f},\tilde{f}^*]
\end{eqnarray}
provided we substitute into eq.(\ref{eq:singlemode})
\begin{equation}
  \label{eq:tildef}
  \tilde{f}(\tau) =\,f(\tau)+i\hbar \frac{\phi}{\lambda_0 }
  \delta^u(\tau)
\end{equation}
and
\begin{equation}
  \label{eq:tildefstar}
  \tilde{f}^*(\tau) =\,f(\tau)-i\hbar \frac{\phi^*}{\lambda_0}
  \delta^{l}(\tau)
\end{equation}                                                            
where $\delta^{u(l)}(\tau)$ is a delta function on the upper (lower)
part of the contour, and vanishes on the lower (upper) part.

The single mode generating functional involves a Gaussian average and
is given by the quadratic form
\begin{equation}
  Z[f,f^*] = e^{-\frac{i}{{\hbar}^2}\int_C d\tau
    \int_C d\tau ' \,\lambda_0^2 \,
    f^*(\tau)\,B(\tau,\tau')\,f(\tau')},
\end{equation}
where $B$ is the single mode Green's function
\begin{equation}
  B(\tau,\tau')=-i\langle 0| T_C(b_0(\tau) b^{\dag}_0(\tau'))| 0\rangle,
\end{equation}
the zero temperature limit of the previously introduced bath
propagator $D(\tau,\tau')$ for the considered mode.  If we therefore
insert the proper "force" $f^{0}$ according to the prescription
eqs.(\ref{eq:tildef},\ref{eq:tildefstar}) we obtain the influence
function for an environment in a coherent state
\begin{eqnarray}
  Z(t_4,t_3,t_2,t_1) & = & Z[f_0] \nonumber \\
  &=& Z_{T=0}(t_4,t_3,t_2,t_1) \nonumber \\
  & & \exp \Bigg[
  2i\frac{\lambda_0}{\hbar \omega_0} \Im m \bigg\{ \phi (e^{-i\omega_0
    t_2} - e^{-i\omega_0 t_1}  \nonumber \\
  &&\hspace{10mm} +e^{-i\omega_0 t_4} -e^{-i\omega_0 t_3}) \bigg\} \Bigg].
\end{eqnarray}
Here $Z_{T=0}$ denotes the influence function for the thermal case
with the temperature set equal to zero, that is, the ground state
case.

Exploiting the observations already made for the external field case
we obtain the following expression for the transmission coefficient
\begin{eqnarray}
  T(\epsilon,\epsilon ') & = & \frac{2\Gamma_l \Gamma_r}{\pi\hbar^3}
  \int_{-\infty}^{\infty} \! d\tau '' \int_0^{\infty} \!\! d\tau '
  \int_0^{\infty} \!\! d\tau\, \tilde{Z}_{T=0}(\tau, \tau ', \tau '')
  \nonumber \\
  & & J_0\left(\frac{2\lambda_0| \phi |}{\hbar \omega_0 }
  | f_{\tau ,\tau' ,\tau ''}^{\omega_0 } | \right) \nonumber \\
  & &\exp\left\{-\frac{i}{\hbar} (\epsilon -\epsilon ') \tau '' -
  \frac{i}{\hbar} (\epsilon ' - \tilde{\epsilon_c} -i \Gamma ) \tau'
  \right. \nonumber \\
  && \hspace{10mm}\left. + \frac{i}{\hbar}
  (\epsilon ' - \tilde{\epsilon_c} + i \Gamma )\tau \right\},
\end{eqnarray}
where the function $f_{\tau ,\tau' ,\tau ''}^{\omega_0}$ is the same as
the function introduced in eq.(\ref{eq:timefunc}).

The coherent state case shares features with both the thermal and the
external field cases. The coherent state influence function consists
of a factor equal to the zero temperature Einstein model
influence function, which describes the systematic influence and
quantum zero point fluctuations, and in addition a factor identical to
the external field influence function.

For the total transmission coefficient we get
\begin{eqnarray}
  T(\epsilon' ) & = & \frac{4 \Gamma_l \Gamma_r}{\hbar^2}
  \int_0^{\infty} \!\! d\tau \int_0^{\infty} \!\! d \tau' \,
  J_0\left(\frac{2\lambda_0|  \phi |} {\hbar \omega_0 }
  | f_{\tau ,\tau' ,\tau ''=0}^{\omega_0 } | \right) \nonumber \\
  &&\exp\left\{ -\left( \frac{\lambda_0}{\hbar \omega_0} \right)^2
  \left( 1 - e^{-i\omega_0 (\tau - \tau')} \right)  \right\} \nonumber \\
  &&\exp\left\{ - \frac{i}{\hbar} (\epsilon ' - \tilde{\epsilon_c} -i
  \Gamma ) \tau ' + \frac{i}{\hbar} (\epsilon ' - \tilde{\epsilon_c}
  + i \Gamma )\tau \right\}.
\end{eqnarray}
Performing a calculation similar to the one for the  oscillatory level
model we get a simple formula for the total transmission coefficient
\begin{equation}
  T(\epsilon ') = 4 \Gamma_l \Gamma_r
  \sum_{n=-\infty}^{\infty}\frac{ Q_n } {(\epsilon' - \tilde{\epsilon}_c
    - n\hbar \omega_0)^{2} + \Gamma^{2}},
\end{equation}
where
\begin{equation}
  Q_n= e^{- \left( \frac{\lambda_0}{\hbar\omega_0} \right)^{2} }
  \sum_{k=0}^{\infty} \frac{1}{k!} \left(
  \frac{\lambda_0}{\hbar\omega_0} \right)^{2k} J_{n-k}^{2} \left(
  \frac{2\lambda_0|\phi|}{\hbar\omega_0} \right) .
\end{equation}
In the limit $\phi = 0$, we of course recover the zero temperature
Einstein model result, eq.(\ref{eq:lineweight_T=0}). In the limit of
$\lambda_0 \rightarrow 0$, $\phi \rightarrow \infty$, and $\lambda_0
|\phi| ={\rm constant}$, we recover the classical oscillatory level
model result, eq.(\ref{eq:exttottrans}). This cross-over behavior is
illustrated in figure \ref{fig:coherent}.

\section{Suppression of Quantum Interference}
\label{sec:destruction}

In the following section we investigate how the different environments
influence the phase coherence in a quantum interference setup. We
envisage the situation where two resonant levels coupled to
environments are placed in parallel and transport can take place
through either\cite{Twolevel}.  A physical realization could be double
barrier structures situated on the two arms of a ring coupled to two
reservoirs. The Hamiltonian still has the form
\begin{equation}
  H= H_e+H_b+H_i,
\end{equation}
but now the sample Hamiltonian corresponds to two levels
\begin{equation}
  H_s=\sum_{c=1,2} \epsilon_c\,a_c^{\dag}a_c.
\end{equation} 
The tunneling can take place through either level with different
couplings
\begin{equation}
  H_t=\sum_{{\bf p},{\sigma},c}\,
  \{V_{{\bf p},{\sigma,c}}\, a_c^{\dag}
  a_{{\bf p},{\sigma}} + {\rm h.c.}\}.
\end{equation}
The interaction takes the form
\begin{equation}
  H_i = \sum_{c=1,2} a_c^{\dag}a_c \,X_c,
\end{equation}
where the environment operator depends on the level through the
coupling constant
\begin{equation}
  X_c = \sum_{\alpha}\lambda_{\alpha,c} \{ b_{\alpha,c}^{\dag} +
  b_{\alpha,c} \},
\end{equation}
and we assume a situation where the double barrier structures are
coupled to separate environments
\begin{equation}
  H_b = \sum_{c=1,2;\alpha } \hbar \omega_{\alpha} \left\{
  b_{\alpha,c}^{\dag} b_{\alpha,c} + \frac{1}{2} \right\}.
\end{equation}

In order to have an external parameter to vary we envisage an
Aharonov-Bohm type situation by piercing the ring with a magnetic
flux, $\Phi$, so that the propagators change according to propagation
around the different arms of the ring according to
\begin{equation}
  G_1^R \mapsto e^{\frac{i}{2}\Phi /\Phi_0} G_1^R ,
  \hspace{5mm} G_2^R \mapsto e^{-\frac{i}{2}\Phi /\Phi_0}
  G_2^R.
\end{equation}
The transmission probability given by eq.(\ref{eq:condprob}) now consists of
transmission through either arm and to accommodate this two-level
situation eq.(\ref{eq:transprob}) is changed into
\begin{eqnarray}
  P_{{\bf p}',l\rightarrow {\bf p}r}(t) & = &
  \frac{1}{\hbar^{4}}\sum_{c_1,c_2,c_3,c_4}
  V_{{\bf p}',l,c_1,} V_{{\bf
          p},r,c_2}^* V_{{\bf p},r,c_3}
  V_{{\bf p}',l,c_4}^* \nonumber \\
  &&\int_0^t dt_1 \int_0^t dt_2 \int_0^t dt_3 \int_0^t dt_4
  \nonumber \\ & & \exp \left\{ \frac{i}{\hbar}
  \epsilon_{{\bf p},r}(t_2-t_3)+
  \frac{i}{\hbar} \epsilon_{{\bf
          p}',l}(t_4-t_1)\right\} \nonumber \\
  &&\langle \hat{G}_{c_4,c_3}^A(t_4,t_3)
  \hat{G}_{c_2,c_1}^R(t_2,t_1)\rangle.
\end{eqnarray}
where
\begin{equation}
  \hat{G}_{c_2,c_1}^R(t_2,t_1) = -i\theta(t_2-t_1) \langle 0 |
  [\hat{a}_{c_2}(t_2),\hat{a}_{c_1}^{\dag}(t_1)] | 0 \rangle
\end{equation}
and
\begin{eqnarray}
  \hat{G}_{c_4,c_3}^A(t_4,t_3)&=&\left[ \hat{G}_{c_3,c_4}^R(t_3,t_4)
  \right]^* \nonumber \\
  &=& i\theta(t_3-t_4) \langle 0
  |[\hat{a}_{c_4}(t_4),\hat{a}_{c_3}^{\dag}(t_3)] | 0 \rangle.
\end{eqnarray}
In the following we shall neglect all terms except those for which
$c_4=c_3$ and $c_2=c_1$. This is justified if the two resonant levels
have an energy difference larger than the width of the levels. Here we
simply implement it corresponding to propagation taking place through
either arm.

The transmission probability given by eq.(\ref{eq:condprob}) then
consists of the transmission probabilities for transmission through
either arm and an interference contribution:
\begin{equation}
  P_{{\bf p}',l\rightarrow {\bf p},r} (t) = \sum_{c=1,2} P_c({\bf
    p},{\bf p'},t) + P_{int}({\bf p},{\bf p'},t),
\end{equation}
where
\begin{eqnarray}
  P_c({\bf p},{\bf p'},t)&=&\frac{|V_{{\bf p}',l,c}|^2 |V_{{\bf
        p},r,c}|^2}{\hbar^{4}} \nonumber \\
  &&\int_0^t dt_1 \int_0^t dt_2 \int_0^t dt_3 \int_0^t dt_4 \nonumber \\
  &&\exp \left\{ \frac{i}{\hbar} \epsilon_{{\bf p},r}(t_2-t_3)+
  \frac{i}{\hbar} \epsilon_{{\bf p}',l}(t_4-t_1)\right\} \nonumber \\ & &
  G_c^A(t_4,t_3) G_c^R(t_2,t_1) Z_c(t_4,t_3,t_2,t_1)
\end{eqnarray}
and the interference term is given by
\begin{eqnarray}  
  P_{int}({\bf p},{\bf p'},t) & = & \frac{1}{\hbar^{4}}
  V_{{\bf p}',l,1} V_{{\bf
          p},r,1}^* V_{{\bf p},r,2}
  V_{{\bf p}',l,2}^* \nonumber \\ & &
  \int_0^t dt_1 \int_0^t dt_2 \int_0^t dt_3 \int_0^t dt_4
  \nonumber \\ & & \exp \left\{ \frac{i}{\hbar}
  \epsilon_{{\bf p},r}(t_2-t_3)+
  \frac{i}{\hbar} \epsilon_{{\bf
          p}',l}(t_4-t_1)\right\} \nonumber \\ & &
  G_2^A(t_4,t_3) G_1^R(t_2,t_1) \nonumber \\ & &
  Z_1(0,0,t_2,t_1)Z_2(t_4,t_3,0,0) + {\rm c.c.} ,
\end{eqnarray}
We now have an influence function for each of the two different
environments, $Z_1$ and $Z_2$, corresponding to the two different arms
of the interferometer. For the transmission coefficient we now have
according to eqs.(\ref{eq:tprobpertime},\ref{eq:transfirst})
\begin{eqnarray}
  T(\epsilon, \epsilon' ) & = & \lim_{t\rightarrow \infty} \frac{1}{t} \;
  \sum_{{\bf p}, {\bf p}' } P_{{\bf p}',l \rightarrow {\bf p},r }(t)
  \delta(\epsilon'-\epsilon_{{\bf p}',l } ) \delta(\epsilon
  -\epsilon_{{\bf p} ,r } ) \nonumber \\
  &=& \sum_{c=1,2} T_C(\epsilon, \epsilon' ) + T_{int}(\epsilon,
  \epsilon').
\end{eqnarray}
The calculation of the total transmission coefficient, specified in
eq.(\ref{eq:tot_trans_coeff}), in various limits is similar to what
has been calculated in section \ref{sec:environments}. For
example, in the absence of an environment we get for the total
transmission coefficient
\begin{eqnarray}
  T(\epsilon') &=& T_1(\epsilon') + T_2(\epsilon') +
  2\sqrt{T_1(\epsilon')T_2(\epsilon')} \nonumber \\ & &
  \times \frac{|\Gamma_{12r}||\Gamma_{21l}|}{\sqrt{\Gamma_{1l} \Gamma_{1r}
      \Gamma_{2l} \Gamma_{2r}}} \nonumber \\ & &
  \times \cos\Bigg( \phi - \arctan \frac{\Gamma_1 }{\epsilon' -
    \epsilon_1 } \nonumber \\ & &
  + \arctan\frac{\Gamma_2 }{\epsilon' - \epsilon_2 } +
  \arg{\Gamma_{12r}\Gamma_{21l}^*} \Bigg),
\end{eqnarray}
where $\phi= 2\pi\Phi/\Phi_0$ is the relative phase difference due to
the external flux, and $T_1$ and $T_2$ are the total transmission
coefficients for the individual arms, and specified in the previous
sections. The decay rates, whose energy dependence can be neglected in
the wide band limit, are defined as
\begin{equation}
  \Gamma_c = \Gamma_{cr} + \Gamma_{cl}
\end{equation}
and
\begin{equation}
  \Gamma_{c_1c_2\sigma}(\epsilon) = \pi \sum_{{\bf p}} V_{{\bf
      p}\sigma c_1}^* V_{{\bf p}\sigma c_2}
  \delta(\epsilon-\epsilon_{{\bf p}\sigma}).
\end{equation}
To illustrate the special features of the suppression of phase
coherence in the present model, we discuss the cases of thermal bath,
fluctuating level, classical field, and coherent state environment
below.

\subsection{Thermal bath and fluctuating level cases}

The transmission coefficient through either arm we obtained
previously. For the interference term, we have the expression
\begin{eqnarray}
  T_{int}(\epsilon,\epsilon') & = &\frac{2}{\pi\hbar^3} \; e^{ i
    \phi} \; \Gamma_{12r}\Gamma_{21l} \nonumber \\ & &
  \int_{-\infty}^{\infty} \! d\tau ''
  \int_0^{\infty} \!\! d\tau' \int_0^{\infty} \!\! d\tau \nonumber \\ & &
  \tilde{Z}_1(\tau,0,0 ) \tilde{Z}_2(0,\tau',0) \nonumber \\ & &
  \exp\left\{ - \frac{i}{\hbar} (\epsilon -\epsilon') \tau ''\right\}
  \nonumber \\
  && \exp\left\{-\frac{i}{\hbar} (\epsilon' - \tilde{\epsilon_2} -i \Gamma_2 )
  \tau '\right\} \nonumber \\
  && \exp\left\{+ \frac{i}{\hbar} (\epsilon' - \tilde{\epsilon_1} + i
  \Gamma_1 )\tau \right\} + {\rm c.c.} ,
\end{eqnarray}
where the influence function corresponding to site $c$ enters as
\begin{eqnarray}
  \tilde{Z}_c(\tau,0,0) &=& \exp\left\{ -\frac{i}{4}\int_0^{\infty}
  \!\! d\omega\, \frac{J_c(\omega)}{\omega^{2}} \sin\omega\tau \right\}
  \nonumber \\
  &&\exp\Bigg\{-\frac{1}{4}\int_0^{\infty} \!\! d\omega\,
  \frac{J_c(\omega)}{\omega^{2}} \coth\frac{\hbar\omega}{2k_{B}T}
  \nonumber \\
  && \hspace{15mm}\times (1-\cos\omega\tau) \Bigg\},
\end{eqnarray}
and the central site energies for each arm are shifted downwards
according to
\begin{equation}
  \tilde{\epsilon}_c=\epsilon_c - \frac{\lambda_c }{4} ,
\end{equation}
with the negative polaronic energy shift
\begin{equation}
  \lambda_c = \int_0^{\infty} \!\! d\omega \,
  \frac{J_c(\omega)}{\omega},
\end{equation}
being site dependent through the spectral function
\begin{equation}
  J_c(\omega) = \frac{4}{\hbar^{2} } \sum_{\alpha }
  \lambda_{\alpha,c}^{2} \delta(\omega - \omega_{\alpha } ).
\end{equation} 
In the course of the derivation we have noted that
\begin{equation}
  \tilde{Z}_c(\tau,0,0) = \tilde{Z}_c^{\ast}(0,\tau,0),
\end{equation}
which follows from eq.(\ref{eq:hermconj}).
The interference contribution to the transmission coefficient is seen
to have the form
\begin{eqnarray}
  \label{eq:int_trans_coeff}
  \lefteqn{T_{int}(\epsilon,\epsilon')} \nonumber \\
  & = &\frac{4}{\hbar^2} \; e^{ i
    \phi } \; \Gamma_{12r}\Gamma_{21l}
  \delta(\epsilon - \epsilon' ) \nonumber \\
  &&\int_0^{\infty} \!\! d\tau' \,
  \tilde{Z}_2(0,\tau',0) \exp\left\{ - \frac{i}{\hbar} (\epsilon' -
  \tilde{\epsilon_2} -i \Gamma_2 ) \tau ' \right\} \nonumber \\
  & & \int_0^{\infty} \!\! d\tau \, \tilde{Z}_1^{\ast}(0,\tau,0)
  \exp\left\{\frac{i}{\hbar} (\epsilon' - \tilde{\epsilon_1} + i
  \Gamma_1 )\tau \right\}  \nonumber \\
  && + {\rm c.c.}
\end{eqnarray}
We notice, that the interference term vanishes unless the energy of
the particle is conserved. Any energy exchange with the thermal
environment thus destroys the interference. In the present model
interference is thus hypersensitive to the presence of the thermal
environment, and in a dramatic fashion displays the equivalence of
dissipation and loss of phase coherence.

We can also calculate the transmission coefficient for the case of
fluctuating levels.  We assume that each site, $c=1$ and $c=2$, is
coupled to separate classical fluctuating environments, represented by
the variables $X_c(t)$, $c=1,2$, respectively. Analogously to the
previous thermal bath case there are contributions to the transmission
coefficient from transmission through either arm, and there is an
interference contribution specified by eq.(\ref{eq:int_trans_coeff}),
except for the thermal influence function being replaced by
\begin{eqnarray}
  \tilde{Z}_c(\tau,0,0) &=& \tilde{Z}_c(0,\tau,0) \nonumber \\
  &=& \exp\left\{-\frac{1}{4 \hbar^{2} } \int_0^{\tau} \!\! dt
  \int_0^{\tau} \!\! dt' \, K_c(t-t') \right\},
\end{eqnarray}
where
\begin{equation}
  K_c(t,t') = \langle (X_c(t)-\langle X_c(t) \rangle)
  (X_c(t')-\langle X_c(t') \rangle)  \rangle.
\end{equation}
As in the thermal case, the interference contribution to the
transmission coefficient is completely suppressed by dissipation.

\subsection{External field and coherent state cases}

In the case where the energy levels change harmonically
in time with the same frequency
\begin{equation}
  X_c(t) = X_c\cos\omega_0t.
\end{equation}
but with different coupling strength, we obtain for the interference
term
\begin{eqnarray}
  T_{int}(\epsilon - \epsilon' ) & = & \frac{2e^{i\phi}}{\pi \hbar^3}
  \Gamma_{12r} \Gamma_{21l}
  \nonumber \\
  &&\int_{-\infty}^{\infty} \! d\tau'' \int_0^{\infty} \!\! d\tau'
  \int_0^{\infty} \!\! d\tau \nonumber \\ & &
  \exp\left\{-\frac{i}{\hbar} (\epsilon
  -\epsilon') \tau'' \right\} \nonumber \\
  &&\exp\left\{-\frac{i}{\hbar} (\epsilon' -
  \epsilon_2 -i \Gamma_2) \tau '\right\} \nonumber \\
  &&\exp\left\{\frac{i}{\hbar}
  (\epsilon' - \epsilon_1 + i \Gamma_1
  )\tau \right\} \nonumber \\ & & \lim_{t\rightarrow
    \infty}\frac{1}{t}\int_0^t \! dt_1\, \exp \Bigg\{ \frac{i}{\hbar}
  (\Re e g^{\omega_0 }_{\tau ,\tau' ,\tau ''} \sin\omega_0t_1
  \nonumber \\
  &&+ \Im m g^{\omega_0}_{\tau ,\tau' ,\tau ''}
  \cos\omega_0 t_1) \Bigg\} +
  c.c. ,
\end{eqnarray}
where we have introduced the function
\begin{eqnarray}
  g^{\omega_0}_{\tau ,\tau' ,\tau ''} &=& \frac{X_1}{\hbar\omega_0} ( 1 -
  e^{i\omega_0 \tau } ) \nonumber \\
  && + \frac{X_2}{\hbar\omega_0}
  (e^{i\omega_0(\tau + \tau'')} - e^{i\omega_0(\tau+\tau''
    -\tau')}).
\end{eqnarray}
Upon performing the integral over $t_1$ we obtain
\begin{eqnarray}
  T_{int} (\epsilon,\epsilon ') & = & \frac{
    2e^{i\phi}}{\pi\hbar^3} \Gamma_{l2r} \Gamma_{21l}
  \nonumber \\
  &&\int_{-\infty}^{\infty} \! d\tau''
  \int_0^{\infty} \!\! d\tau' \int_0^{\infty} \!\! d\tau\, J_0 \left( |
  g^{\omega_0 }_{\tau ,\tau' ,\tau ''} | \right) \nonumber \\ & &
  \exp\left\{ -\frac{i}{\hbar} (\epsilon -\epsilon ') \tau '' -
  \frac{i}{\hbar} (\epsilon ' - \epsilon_2 -i \Gamma_2 ) \tau ' \right\}
  \nonumber \\
  &&\exp\left\{\frac{i}{\hbar} (\epsilon ' - \epsilon_1  + i \Gamma_1
  )\tau \right\} \nonumber \\
  && + {\rm c.c.}
\end{eqnarray}
Change in energy of the particle due to interaction with the external
field does not destroy the interference in accordance with the notion
that such a non-dissipative process does not lead to suppression of
phase coherence when time reversal symmetry is unbroken.

Carrying out a similar calculation for the coherent state case, we
find a similar qualitative behavior for the interference term.
Exchange of one quantum does not change the coherent state and
interference is not destroyed.

\section{Summary and conclusion}
\label{sec:summary}

We have developed a functional method to study the influence of
various environments on quantum tunneling, and shown that the effect
of the environment on the transmission probability, specified by the
influence function, is described by the value of a certain generating
functional.  The state of the environment only occurs explicitly in
the expression for the influence function through the appearance of
its statistical operator, thereby allowing a unified discussion of the
influence of environments on phase coherence and thereby on the
transmission properties. An advantage of the approach is that
different environments are treated on an equal footing, thereby
simplifying comparison of similarities and differences.

We have calculated the influence function for various environments and
parameter regimes using the developed non-equilibrium generating
functional technique. In order to obtain analytical results we have
concentrated on simple resonant tunneling, and invoked the wide band
approximation.  In the thermal case we obtained analytical results for
the broad resonance limit, the strong coupling limit, the high
temperature limit and the weak coupling and low temperature limit. The
short time approximation used in the broad resonance, strong coupling
and high temperature limits, and the long time approximation invoked in
the weak coupling and low temperature limit were numerically
demonstrated to be good approximations. For the thermal case we
discussed the closed expression obtainable for the Einstein model.

We studied the effect of a classical fluctuating environment on the
tunneling process, the fluctuating level model. The environment is
here described by a Gaussian distributed stochastic variable. The
fluctuating level model was shown to only capture the correlation part
of the influence of the thermal environment, as the fluctuating level
model neglects the systematic part except for its average influence.
We find in this case that the total transmission coefficient is a
Lorentzian with its width temperature broadened.

We studied the effect of a harmonically varying external field, and
found the heights and widths of the Lorentzian sidebands at all the
harmonics of the classical force.

For the coherent state case we also obtain a closed expression for the
influence function, and we showed that the coherent state case is an
intermediate situation sharing features of both the zero temperature
Einstein case and the external field case.

Finally we studied the suppression of quantum interference by
considering a simple model of two resonant levels situated on the two
arms of a ring connected to two external reservoirs. The quantum
interference was monitored by having the ring enclosing a magnetic
flux.  When calculating the transmission probability we obtained
contributions corresponding to tunneling through either arm and an
interference contribution. We found that the interference term
vanishes in the thermal bath and fluctuating level cases even for the
slightest energy exchange between system and environment. The model
thus represents a case where dissipation completely destroys quantum
interference. In the case of a non-dissipative environment, e.g., a
classical force, there is no loss of phase coherence, and there is
therefore always an interference contribution, even if energy is not
conserved in the transition.

In conclusion, we have demonstrated the effiency of the developed
non-equilibrium generating functional technique for evaluating the
environmental influence on coherent quantum processess.

\section*{Acknowledgments}

We (PA, JR) acknowledge stimulating discussions with Karl Hess and Tony
Leggett. This work was supported by the Swedish Natural Research
Council through contracts F-AA/FU 11084-302 (PA) and F-AA/FU 10199-306
(JR).

\begin{figure}[htbp]
  \begin{center}
    \leavevmode
    \caption{The Feynman diagrammatic representation of the decoupling of
      particle and environment dynamics for the amplitude correlator
      $\langle \hat{G}^A_c(t_4,t_3) \hat{G}^R_c(t_2,t_1) \rangle$ in
      the wide band approximation. The double lines denote the full
      central site Green's functions while the single lines denote the
      central site Green's functions in the absence of coupling to the
      environment.  The curly lines denote the correlator of the
      environment operator.}
    \label{fig:decoupling}
  \end{center}
\end{figure}
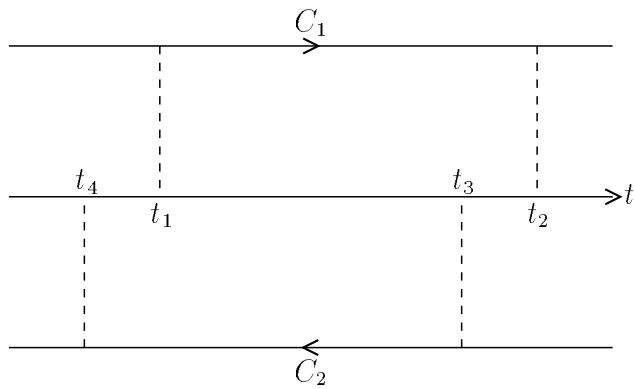
\begin{figure}[htbp]
  \begin{center}
    \leavevmode
    \caption{Closed time path $C$ extending from $-\infty$ to
      $+\infty$ along $C_1$ and back again along $C_2$.}
    \label{fig:path}
  \end{center}
\end{figure}
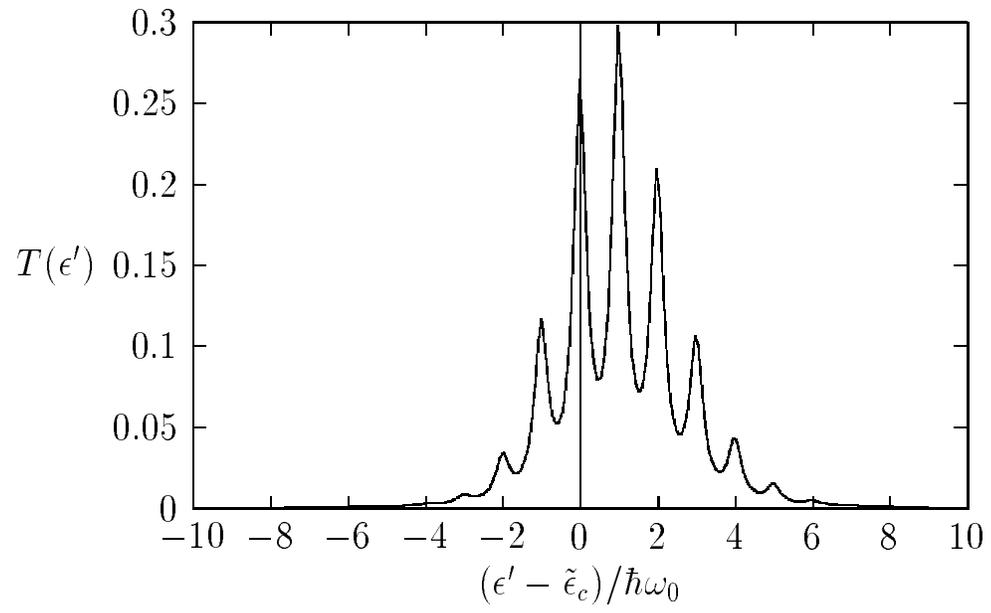
\begin{figure}[htbp]
  \begin{center}
    \leavevmode
    \caption{The total transmission coefficient (in units of
      $4\Gamma_l \Gamma_r/ \Gamma^2$) for the Einstein environment
      case. The parameters are $\Gamma=0.2 \hbar\omega_0$,
      $\lambda_0=\hbar\omega_0$, and $k_BT=\hbar \omega_0$.}
    \label{fig:einstein}
  \end{center}
\end{figure}
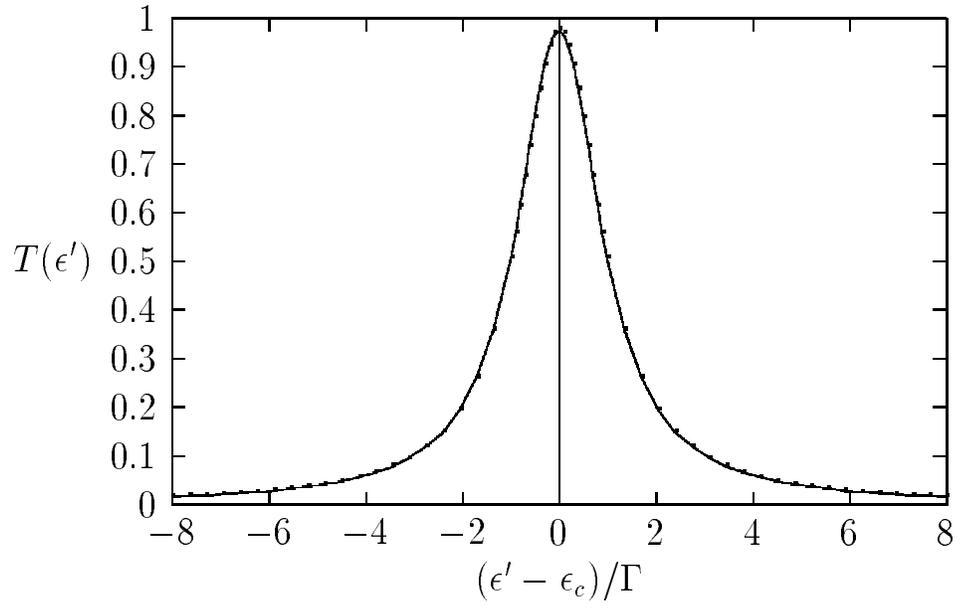
\begin{figure}[htbp]
  \begin{center}
    \leavevmode
    \caption{Transmission coefficient (in units of $4\Gamma_l
      \Gamma_r/ \Gamma^2$) for the ohmic environment case in the broad
      resonance limit, $\Gamma > \hbar \omega_c$. In the figure the
      broad resonance approximation (solid curve) is compared to the
      exact result (dots). The parameter choice is: $\eta=1$,
      $2k_BT=\Gamma$ and $\hbar\omega_c=0.1\Gamma$, which yields
      $\hbar^2 \kappa(T)=0.101 \Gamma^2$.}
    \label{fig:broad}
  \end{center}
\end{figure}
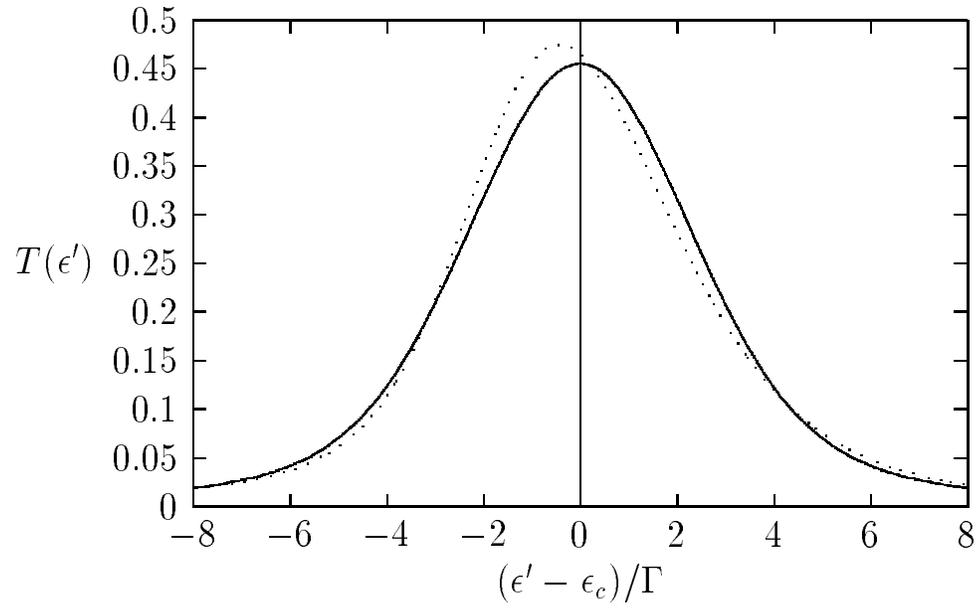
\begin{figure}[htbp]
  \begin{center}
    \leavevmode
    \caption{Transmission coefficient (in units of $4\Gamma_l
      \Gamma_r/ \Gamma^2$) for the ohmic environment case in the
      strong coupling limit, $\eta > 1$. In the figure the strong
      coupling approximation (solid curve) is compared to the exact
      result (dots). The parameter choice is: $\eta=10$, $\hbar
      \omega_c=\Gamma$, $2k_BT=\Gamma$ and $\hbar^2 \kappa(T) \approx
      14.27 \ \Gamma^2$.}
    \label{fig:strong_coupling}
  \end{center}
\end{figure}
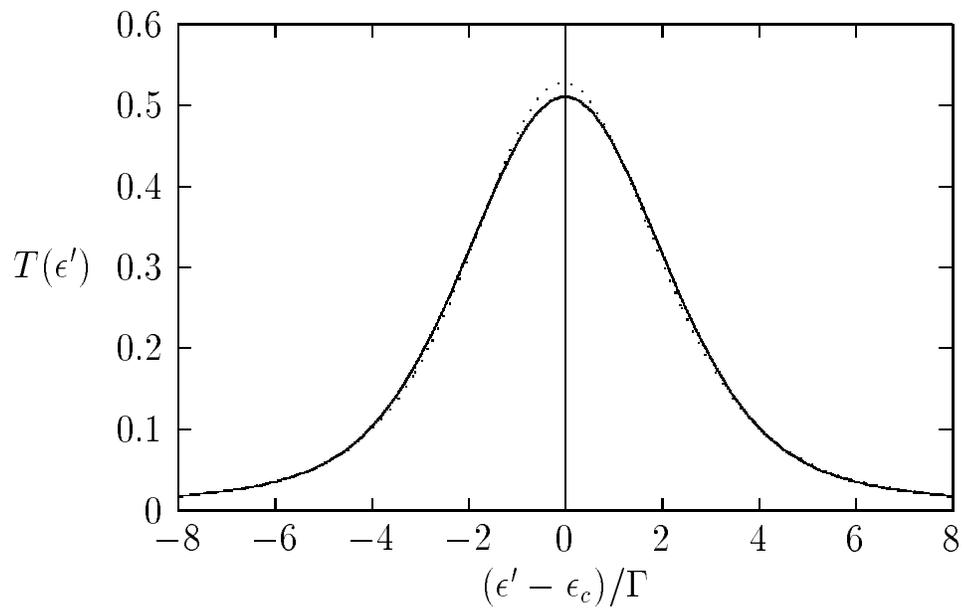
\begin{figure}[htbp]
  \begin{center}
    \leavevmode
    \caption{Transmission coefficient (in units of
      $4\Gamma_l \Gamma_r/\Gamma^2$) for the ohmic environment case in
      the high temperature limit, $k_BT > (1+1/\eta)\hbar \omega_c$.
      In the figure the high temperature approximation (solid curve)
      is compared to the exact result (dots). The parameter choice is:
      $\eta=1$, $\hbar \omega_c=\Gamma$, $2k_BT=10 \Gamma$ and
      $\hbar^2 \kappa'(T)=10 \ \Gamma^2$.}
    \label{fig:high_temp}
  \end{center}
\end{figure}
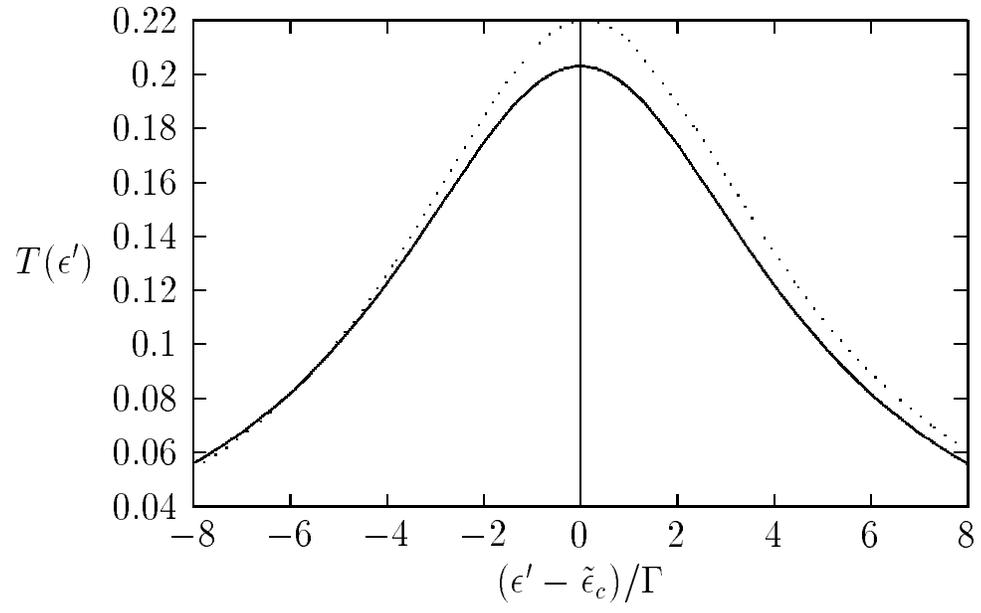
\begin{figure}[htbp]
  \begin{center}
    \leavevmode
    \caption{Transmission coefficient (in units of $4\Gamma_l
      \Gamma_r/ \Gamma^2$) for the ohmic environment case in the weak
      coupling and low temperature limit, $\eta<1$ and $k_BT <
      \frac{\hbar \omega_c}{\eta}$. In the figure the weak coupling
      and low temperature approximation (solid curve) is compared to
      the exact result (dots). The parameter choice is: $\eta=0.1$,
      $\hbar \omega_c=100 \Gamma$ and $2k_BT=100 \Gamma$.}
    \label{fig:wcltl2}
  \end{center}
\end{figure}
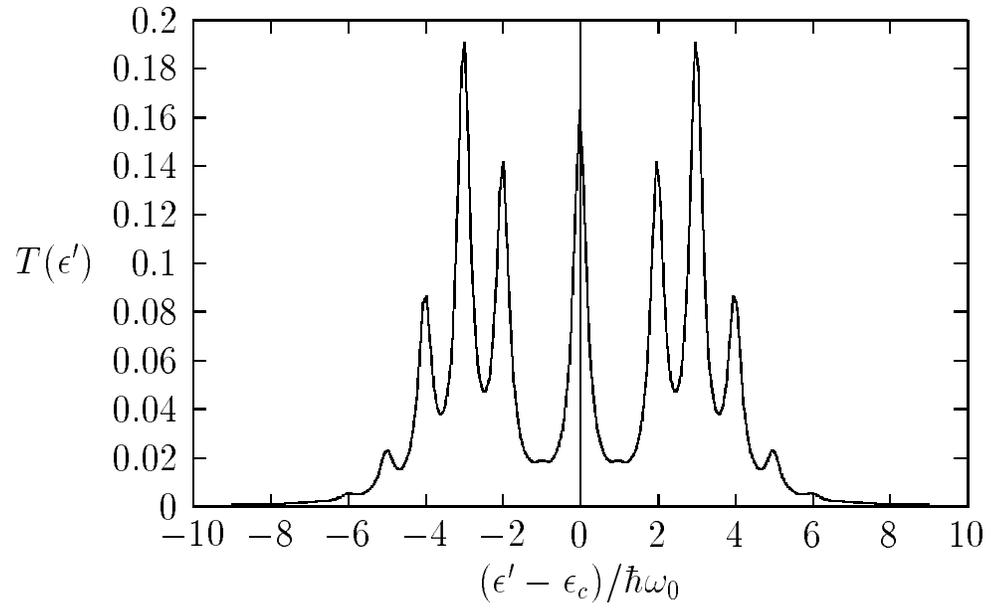
\begin{figure}[htbp]
  \center
  \leavevmode
  \caption{Transmission coefficient (in units
    of $4\Gamma_l \Gamma_r/ \Gamma^2$) for the external field case,
    $X(t)=X_0 \cos \omega_0t$. The parameter choice is:
    $\Gamma=0.2\hbar\omega_0$ and $X_0=4\hbar\omega_0$.}
  \label{fig:external}
\end{figure}
\begin{figure}[htbp]
  \begin{center}
    \leavevmode
    \vspace{4mm}
    \caption{Transmission coefficient (in units of $4\Gamma_l
      \Gamma_r/ \Gamma^2$) for the coherent state environment. In
      diagram (a) we have chosen $\Gamma=0.2\hbar\omega_0$,
      $\lambda_0=\hbar\omega_0$ and $|\phi|=1$. In this case the
      transmission curve resembles the one obtained for the Einstein
      model, see figure \ref{fig:einstein}. In (b)
      $\Gamma=0.2\hbar\omega_0$, $\lambda_0=\hbar\omega_0$ and
      $|\phi|=2$. In (c) $\Gamma=0.2\hbar\omega_0$, $\lambda_0=0.2
      \hbar\omega_0$ and $|\phi|=10$. The shape of the curve in
      diagram (c) resembles the one obtained in the external field
      case, see figure \ref{fig:external}.}
    \label{fig:coherent}
  \end{center}
\end{figure}
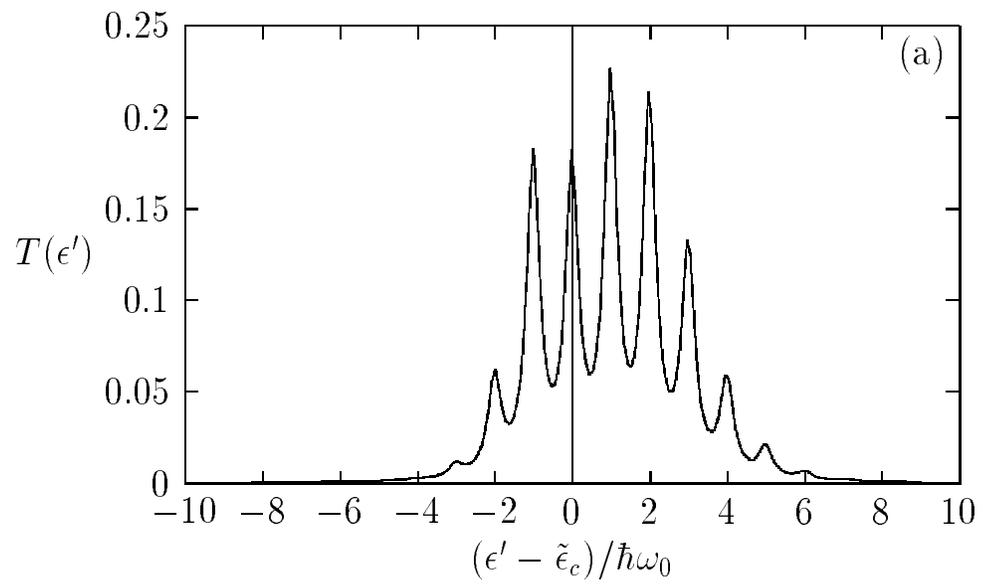
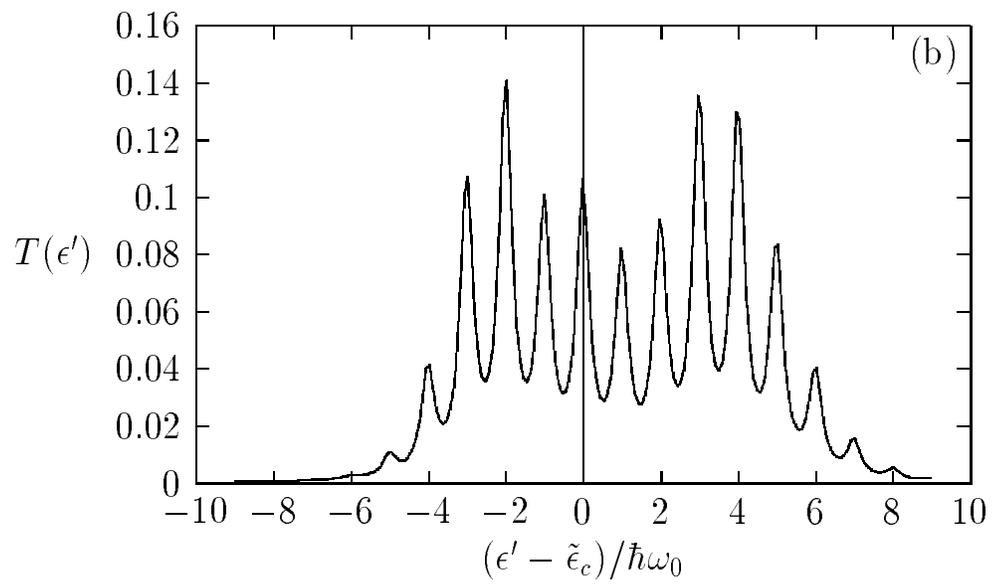
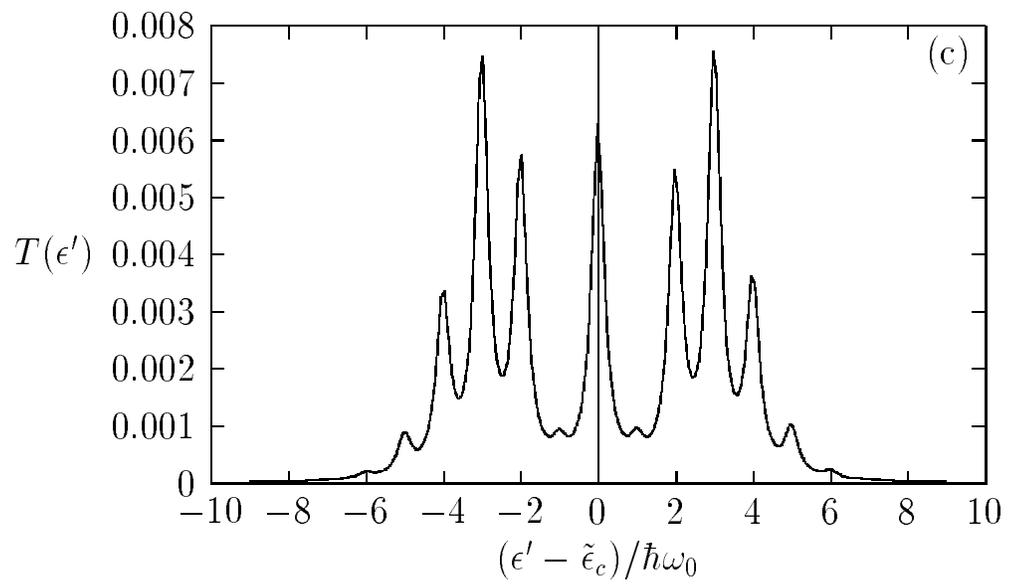
\end{document}